\documentclass{aa}  

\usepackage{graphicx}
\usepackage{txfonts}

\usepackage[colorlinks=true, linkcolor=blue, citecolor=blue, urlcolor=blue, breaklinks=true]{hyperref}
\usepackage{appendix}

\usepackage{xcolor}
\usepackage[normalem]{ulem}

\begin{document}

   \title{Dynamically close galaxy pairs from the unWISE survey}

   \subtitle{Testing the merger-AGN-star formation connection}

   \author{Josephine Chishala
          \inst{1,2},
          Roberto De Propris\inst{2,3}, and 
          Mirjana Povi\'c\inst{4,5,6}
          }

   \institute{Department of Physics, School of Mathematics and Natural Sciences, Copperbelt University, P.O. Box 21692, Kitwe, Zambia.
   \and
   Department of Physics and Astronomy, Botswana International University of Science and Technology, Private Bag 16, Palapye, Botswana.\\
              \email{josephinechishala2@gmail.com}
        \and
             Finnish Centre for Astronomy with ESO, University of Turku, Vesilinnantie 5, Turku, 20014, Finland.
        \and 
             Instituto de Astrofísica de Andalucía (IAA), Spanish National Research Council (CSIC), 18008, Granada, Spain.
        \and
            Department of Astronomy and Astrophysics, Space Science and Geospatial Institute (SSGI), P.O. Box 33679, Addis Ababa,   Ethiopia.
         \and
         Department of Physics, Faculty of Science, Mbarara University of Science and Technology (MUST), P.O. Box 1410, Mbarara, Uganda.}

   \date{}

\abstract   
{Galaxy mergers are expected to have a profound influence on the star formation histories of galaxies. It is generally expected that mergers are the main drivers of galaxy mass growth through the accretion of mass and the triggering of new star formation episodes, while the shocks and torques induced by the merger may drive gas and dust to central supermassive black holes and fuel active galactic nuclei (AGN) activity and producing both positive and negative feedback.}
{We test whether a merger-AGN-star formation connection exists by selecting samples of galaxy pairs of stellar masses $\log(M/M_{\odot}$) $\sim 10.2$ and $\sim 11.4$ within redshift z \(<\) 0.25 at various projected separation and velocity differences in an increasing order, and therefore having a decreasing probability of being truly bound and interacting.}
{We identify galaxies in close pairs and then measure their star formation rates (SFRs) (via their $NUV-r$ colours) and the degree of AGN activity (from X-rays, radio emission at 20cm, WISE infrared colours, and emission line ratios) as a function of their projected separation and velocity difference.}
{We find only weak evidence that galaxies in pairs
have higher SFRs as galaxies become closer in projected and velocity separation, except possibly for pairs at closest separation of rp \(<\) 20 kpc and $\Delta$V \(<\) 500 km/s. Similarly, we see no strong evidence that AGN are more common for galaxies in closer pairs, irrespective of the method used to detect AGN.}
{For this sample, we do not find any clear evidence that mergers and interactions may play a significant role in triggering star formation and AGN activity, opposite to expectations from theoretical models invoking feedback episodes. Secular processes may be more important, although this may depend on the selection of galaxies and indicators for star formation and AGN activity.}


   \maketitle

\section{Introduction}

\begin{figure*}[h!]
    \centering
    \includegraphics[width=1\linewidth]{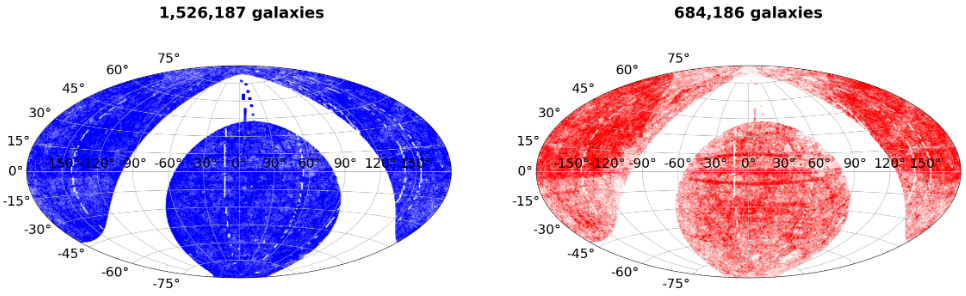}
    \caption{Distribution of all galaxies (left) and galaxies with redshift (right) on the sky.}
    \label{skydist}
\end{figure*}

Within the commonly accepted $\Lambda$ cold dark matter (CDM) model, galaxies are expected to form and assemble hierarchically
\citep[e.g.,][]{White1978,Springel2001,Moreno2013,
Duncan2019}. Galaxy mergers are therefore expected to be a common occurrence. In cosmological hydrodynamical simulations, about 10\% of the stellar mass in a Milky Way-sized galaxy today is accreted from other galaxies, and the fraction of accreted stellar mass increases to $\sim 60\%$ for halos that are 10 times more massive \citep{Pillepich2018}. In the local universe, a few percent of galaxies are observed to be interacting with other companions of similar mass \citep[e.g.,][]{Lin2008,Darg2010,Lotz2011,Xu2012,Lopez2015}. Mergers are therefore a key element in the process of galaxy formation and evolution.

Mergers not only contribute to the growth of stellar mass: interactions can alter galaxy structures, leading to the formation of more bulge-dominated systems \citep{Toomre1972,Barnes1996}, but also potentially creating large disk galaxies in gas-rich scenarios \citep{Robertson2006}.
Mergers can enhance star formation \citep{Scudder2012,Patton2013,Violino2018,Bickley2022,Garay2023} or act as a quenching mechanism \citep{Springel2005,Hopkins2006,Ellison2022,Wilkinson2022}. 
Observations show that the SFRs in close pairs are enhanced relative to a control sample of more isolated
galaxies (e.g.,\citealt{Kartaltepe2012,Scudder2012,Patton2013,Robotham2014,Barrera2015,Pan2018,
Garduno2021,Steffen2021,Shah2022}, among others). Most of the star formation enhancement occurs at close separations 
($ < 30$ $h^{-1}$ kpc), but can be found even at distances approaching 150 kpc \citep{Patton2013}. Numerical simulations bear this out
\citep{Mihos1996,DiMatteo2005,Hopkins2008,Torrey2012,Renaud2014,Moreno2019}, with \cite{Patton2020} recovering this behaviour even in the more recent IllustrisTNG dataset.

Similarly, galaxy merging could play an important role in triggering AGN activity (e.g., \citealt{Ellison2011,Satyapal2014,Ellison2019, Mahoro2019,Bickley2023,Bickley2024,Byrne-Mamahit2023,Byrne-Mamahit2024}) and these can also act to boost and/or quench star formation as well (e.g., see \citealt{Zubovas2017,Trussler2020}). However, there is still a lack of agreement in the previous results. Some studies have found no correlation between galaxy mergers and AGN triggering \citep{Cisternas2011,Villforth2017,Marian2019,Silva2021} while others have found that AGN are statistically more likely to be in merging galaxies \citep{Ellison2011,Ellison2013,Lackner2014,Satyapal2014,Weston2017,Goulding2018}. In addition, several works report an increase in the fraction of AGN in the population of merging galaxies for decreasing nuclear separation below 100\,kpc (e.g, \citealt{
Ellison2011,Silverman2011,
Satyapal2014}), while others do not (e.g, \citealt{Shah2020} for an intermediate redshift sample and \citealt{Jin2021} for local galaxies within MANGA).

If a galaxy is going to merge, it needs a companion to merge with. Hence, pairs of galaxies in close proximity can be used
as a proxy for the future merger rate, although the actual merger fraction and timescales require considerable theoretical
effort to simulate. An alternate approach uses the degree of morphological disturbance, but this identifies either merger
remnants or objects that are already in a high stage of coalescence: see \cite{DePropris2007} and \cite{Desmons2023} for a discussion
and comparison of these two methods on the same samples. Both of these methods have biases and may affect results. Additionally, the number of galaxy pairs in most surveys is generally relatively small, and it is often difficult to obtain statistically significant samples of objects with full photometric and spectroscopic coverage. Similarly, star formation and AGN indicators are not always available. Selection effects, such as disentangling close pairs, redshift incompleteness, the photometric bands used for selection, among others, can also affect the results.

In this paper, we will concentrate on the galaxy pairs with the closest separation as a proxy of interacting galaxies and possible mergers, where a fraction of closer pairs (typically $r_p < 20$ kpc and $\Delta V < 500$ km s$^{-1}$) are expected to merge within short timescales. Here we use a large sample of low spectroscopic redshift galaxy pairs at $z < 0.25$ covering the whole sky outside of the Galactic plane, selected at 3.4\,$\mu$m and with ample redshift coverage. Above $\delta=-30^{\circ}$ we have Charge-Coupled Device (CCD) photometry
from the Panoramic Survey Telescope and Rapid Response System (Pan-STARRS) survey \citep{Chambers2016}, while at lower declinations we use imaging from other photometric surveys including the Two Micron All-Sky Survey (2MASS), the Wide Field Infrared Survey Explorer (WISE), and SkyMapper. We identify galaxy pairs at various projected distances and velocity separations and then use archival Near Ultraviolet (NUV) photometry from Galaxy Evolution Explorer (GALEX) to measure star formation \citep{Morrissey2005,Morrissey2007}. To select AGN, we also use X-ray data from eROSITA \citep{Sunyaev2021,merloni2024srg}, radio data from the NRAO VLA Sky Survey (NVSS) \citep{Condon1998}, mid-infrared (mid-IR) data from the unWISE survey \citep{Lang2014,Schlafly2019} and the Sloan Digital Sky Surveys (SDSS) spectroscopic indices where available \citep{Baldwin1981}.

We present our dataset and galaxy pair selection technique in Section \ref{data}. We show the dependence of SFRs on projected separations of the galaxy pairs, as measured by the $NUV-r$ colour in Section \ref{sec_SF}, and the AGN fraction in Section \ref{sec_AGN}. Results are discussed in Section \ref{discussion}, and our conclusions are presented
in Section \ref{conclusion}. We adopt the latest cosmological parameters from \cite{Aghanim2020}.

\section{Data}
\label{data}
Our photometric dataset comes from the unWISE survey \citep{Lang2014,Schlafly2019} in the W1 (3.4 $\mu$m) band. This has several advantages over previous datasets: it covers the entire sky, and, at this wavelength, foreground extinction is nearly negligible. Furthermore, the 3.4$\mu$m band of WISE is dominated by the light from old stars and can be used as an effective measure of stellar mass; by selecting galaxies in this bandpass, we are essentially selecting them by stellar mass \citep{Wen2013,Cluver2014}, with minimal influence from young stars. We select all objects classified as galaxies in the unWISE catalogue ("spread model" $> 0.003$) with W1 $ < 14.7$ (Vega magnitudes), where the unWISE catalogue is stated to be 100\% complete \citep{Schlafly2019}. We also restrict our sample to objects with $| b | > 20$ to avoid the crowded (and highly extincted) regions near the galactic plane. Taking all this into account, we are left with 1,526,187 galaxies in total as our photometric sample. Figure~\ref{skydist} (left plot) shows the distribution of all galaxies on the sky.
\begin{figure}
    \centering
    \includegraphics[width=1\linewidth]{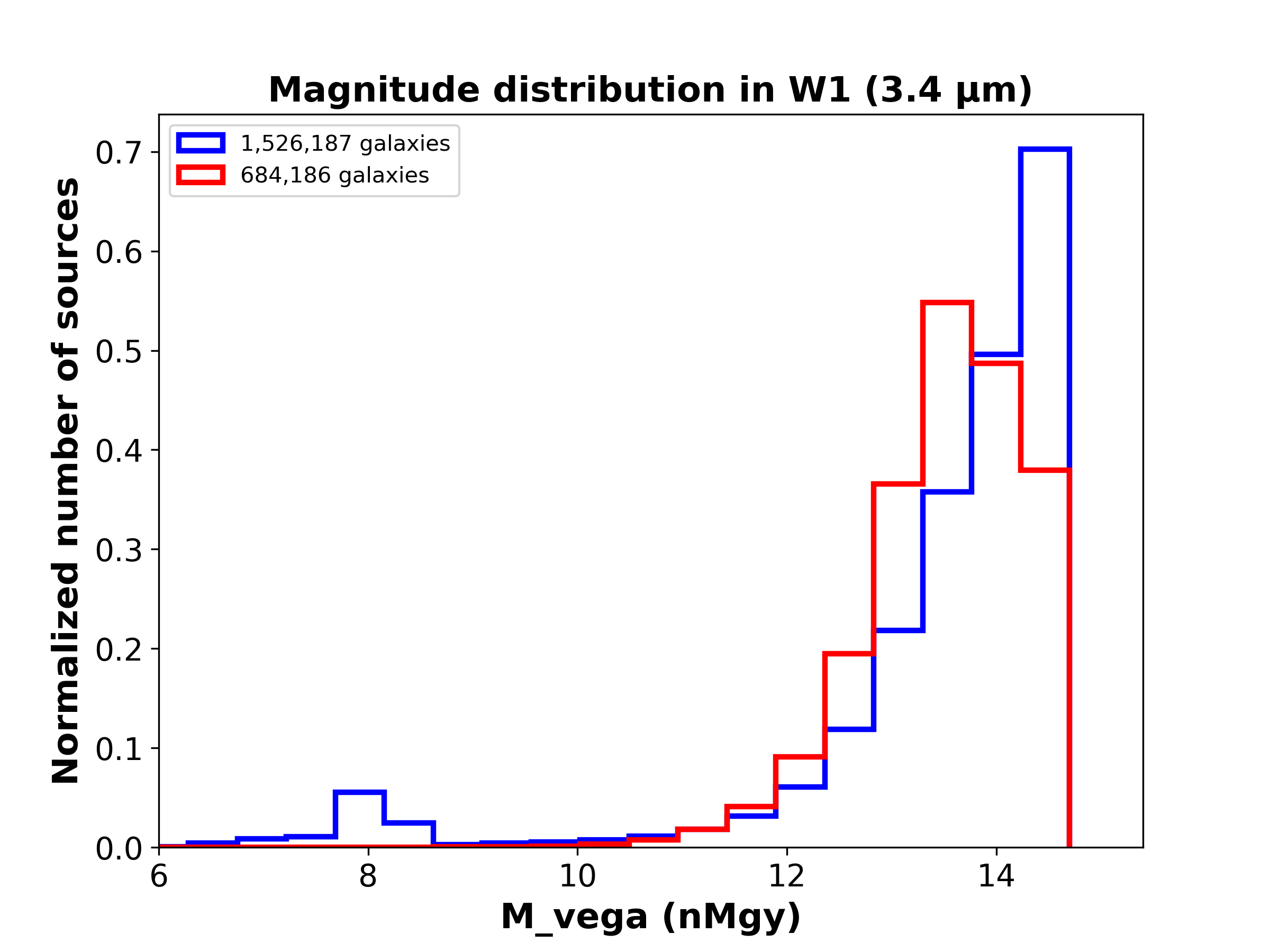}
    \caption{WISE W1 magnitude comparison of the total photometric sample (blue) and the sample with spectroscopic redshifts (red), normalised by the area under the curve.}
    \label{fig:enter-label}
\end{figure}
\begin{figure}
    \centering
    \includegraphics[width=1\linewidth]{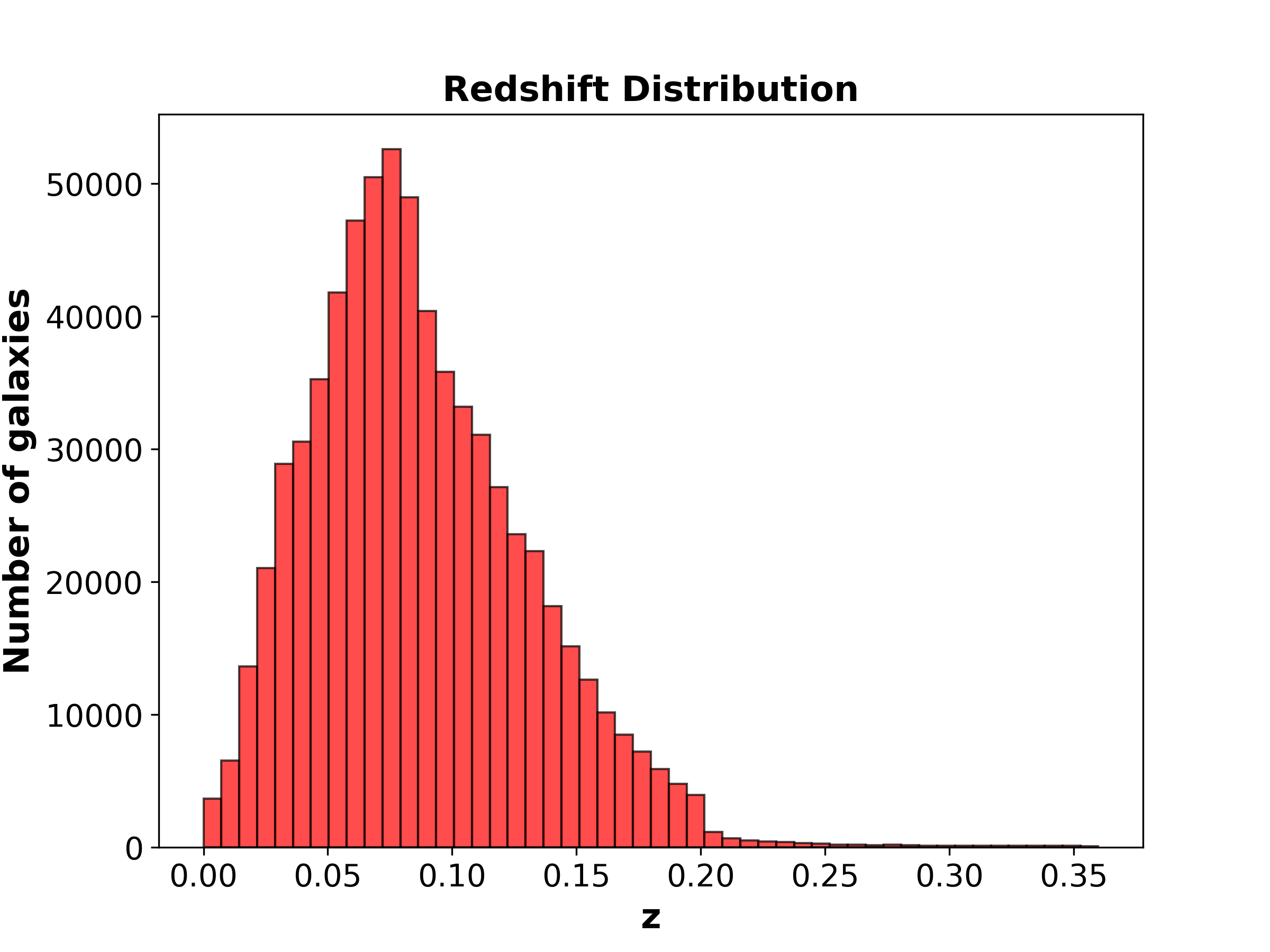}
    \caption{Spectroscopic redshift distribution of the 684,186 selected galaxies.}
    \label{fig:placeholder}
\end{figure}

We then match our galaxies with the NASA/IPAC Extragalactic Database Local Volume Sample (NED-LVS)\footnote{The NASA/IPAC Extragalactic Database (NED) is operated by the Jet Propulsion Laboratory, California Institute of Technology, under contract with the National Aeronautics and Space Administration (http://ned.ipac.caltech.edu).} from \cite{Cook2023} using only the collected spectroscopic redshifts for extragalactic sources from existing surveys. Typical errors in redshift are of the order of a few km s$^{-1}$ and therefore much smaller than the velocity intervals we use in this paper.
We have experimented with several matching radii
between the photometric and spectroscopic samples and find that
the best compromise between minimising the number of lost counterparts and the number of spurious matches is a matching
radius of $5''$. This results in a total of 684,186 galaxies (45\% of the total photometric sample) with spectroscopic redshifts from the 
NED-LVS. Figure~\ref{skydist} (right plot) shows the distribution of sources with available redshifts. Fig.~\ref{fig:enter-label} shows the comparison in $M_{W1}$ magnitude (not dust and K-corrected due to small values of these quantities, where the W1 band may be considered a close proxy of the galaxy's stellar mass \citep[see e.g.,][]{Cluver2014}) between the photometric (in blue) and redshift (in red) samples.

\subsection{Galaxy Pairs}

\begin{figure}[h!]
   \centering
    \includegraphics[width=0.9\linewidth]{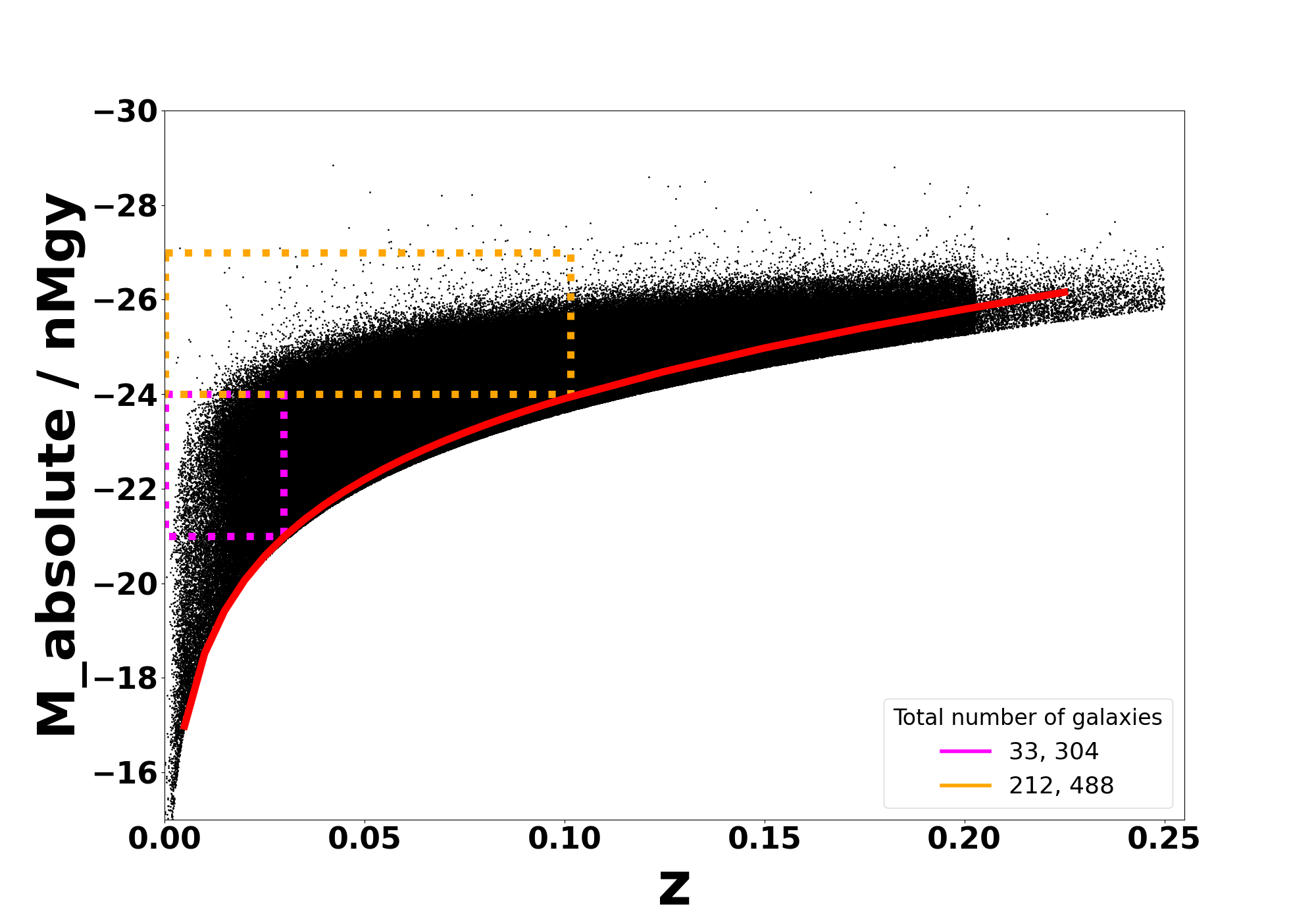}
    \caption{$M_{W1}$ vs. spectroscopic redshift for galaxies in our 
    sample. The red line represents the predicted evolution
    for a slowly evolving galaxy with $W1=14.7$ at $z=0$. The violet and orange boxes define the two volume-limited samples we search for pairs in. We refer to the violet box ($-21 < M_{W1} < -24$ and z \(<\) 0.03) as Box 1 and to the orange box ($-24 < M_{W1} < -27$ and z \(<\) 0.10) as Box 2.}
    \label{boxes}
\end{figure}

Following the methods detailed in \cite{Patton2000,Patton2002}, \cite{DePropris2005,DePropris2007}, and several successive studies, we select galaxies in two volume-limited boxes with $-21 < M_{W1} < -24$ at z $<$ 0.03 (hereafter Box 1), and $-24 < M_{W1} < -27$ at z $<$ 0.10 (hereafter Box 2). Fig.~\ref{boxes} shows the diagram of $M_{W1}$ versus redshift, with galaxies selected in Box 1 and Box 2. The thick red line shows the predicted redshift evolution of a galaxy with W1 magnitude of 14.7 at
$z=0$ assuming a \cite{Bruzual2003} model with a formation redshift of 3 and e-folding time for star formation of 0.3 Gyr. This is typical of a slowly evolving early-type galaxy and therefore the slowest evolving object, which provides the magnitude limit of the survey as a function of redshift. Within these boxes we identify pairs of galaxies
as a function of projected separation and velocity difference, in order to select objects that (in the closest separations) are likely to merge within short timescales (at least for a fraction of truly bound systems). The typical luminosity ratio (which is close to the mass ratio in W1) of these pairs is 1:3. Box 1 contains 33,304 galaxies while Box 2 contains 212,488 galaxies. Using the calibration by \cite{Wen2013} and \cite{Cluver2014} the central absolute W1 luminosities of $-22.5$ (Box 1) and $-25.5$ (Box 2)
correspond to stellar masses of $\log(M/M_{\odot}$) $\sim 10.2$ and $\sim 11.4$, respectively. The magnitude ranges used for these two volume-limited boxes correspond to typical mass ratios of $\sim$ 1:4 \citep{Wen2013,Cluver2014}. By design, the two samples in Box 1 and 2 do not overlap and are independent of each other, and therefore, we have two samples of galaxies where we search for pairs of galaxies in two separate luminosity ranges. The typical standard deviation of W1 derived masses when compared with Galaxy And Mass Assembly (GAMA) stellar masses is about 0.5 \citep{Cluver2014}, while the GAMA and W1 derived stellar masses are in good agreement.

Within these volume-limited samples defined by these two boxes, we search for galaxy pairs where galaxies have projected separation of
$r_p$ $< 20$ kpc, $20 < r_p < 50$ kpc, $50 <r_p < 100$ kpc, $100 < r_p < 250$ kpc, and $500 < r_p < 1000$ kpc with velocity differences 
between pair members of $< 500$ km s$^{-1}$ or 500-1000 km s$^{-1}$. Only unique galaxy pairs (with closest galaxy members) are selected in this construction.

This selects galaxies in pairs that are increasingly more likely to be undergoing interactions and/or to merge as a function of decreasing projected and velocity separations. According to \citep{DePropris2007, DePropris2014} and \citep{Desmons2023}, at least 35\% of pairs within projected separation of rp $<$ 20\,kpc are likely to be
interacting and merging, while \cite{patton2016galaxy} showed that these interactions may occur up to rp $<$ 50\,kpc. Therefore, in this work we consider galaxies in pairs with rp $<$ 20\,kpc and velocity difference of $< 500$ km s$^{-1}$ to be a proxy of merging galaxies. On the other hand,
more widely separated galaxies favour more isolated objects that are less likely to merge or interact:  e.g., the M31/Milky Way pair may be taken as an example of a widely separated pair that may (\citealt{Cox2008}) or may not (\citealt{Sawala2025}) merge only within
a few Gyr\footnote{This also provides an example of how even close pairs may suffer contamination, as from some angles the M31/Milky Way system would appear as having small projected separation such as that they would merge within 1 Gyr.}. Previous studies \citep[e.g.,][]{DePropris2007,DePropris2014,patton2016galaxy,Desmons2023} show that galaxies separated by more than 100--160 kpc $h^{-1}$ do not show clear signs of morphological disturbance and are, therefore, much less likely to merge or interact in the near future.

Our more distant pairs (i.e., in the more distant bins on projected separation and velocity difference between the pair members) may be taken as 'control' samples of more isolated galaxies. A projected separation of 1 Mpc is equivalent to about 1/2 of $r_{200}$ for typical nearby clusters, while a velocity difference of 1000 km s$^{-1}$ is as large as their typical velocity dispersion. \cite{Patton2013} defines galaxies as isolated (for a comparison sample) if they have no bright companion within a projected distance of 2 Mpc and a velocity difference of 1000 km s$^{-1}$, while \cite{Argudo2015} show that galaxies with $|\Delta V| > 200$ km s$^{-1}$ and projected separation $\gtrsim 500$ kpc are very likely to be related to the underlying large-scale distribution of galaxies and therefore not interacting with each other.

We use these samples to derive the
dependence of star formation and AGN activity on the projected separation and velocity difference of galaxies in pairs. Table~\ref{table:1} shows the number of pairs in each absolute magnitude (stellar mass) range, and separation and velocity difference subset. The total number of identified pairs in Box 1 and 2 is 12122 and 39757, respectively, with the total number of galaxies in pairs of 24244 and 79514, respectively. Note that each galaxy is only paired once.\\

\indent The fraction of close pairs at $r_p < 20$ kpc $h^{-1}$ and $\Delta V < 500$ km s$^{-1}$ (that may be expected to merge within $\sim 1$ Gyr) is approximately 2.9\% for Box 1 and approximately 1\% for Box 2. These values are broadly consistent with previous studies, despite possible incompleteness for closer pairs. Redshifts for pair galaxies may not be complete, especially for the closest objects, because of fibre or slit collisions. On the other hand, our redshifts come from numerous independent surveys, which should mitigate this effect. However, our interest here is in comparing the relative fractions of star-forming and AGN hosting galaxies in pairs as a function of pair members' separation. As long as redshift incompleteness for closer pairs does not depend on the measured $NUV$ colour, or measure of AGN activity we choose, the relative fractions of galaxies in pairs identified as star-forming, or showing signs of AGN activity, should not be affected.

\begin{table*}[]
   \centering
       \caption{Number of galaxy pairs as a function of projected separation, velocity difference, and absolute magnitude range}.
    \label{table:1}
    \begin{tabular}{lcccc}
  \hline
      $r_p$ & Number of pairs & Number of pairs & Number of pairs & Number of pairs \\
       & \textbf{Box 1 }&\textbf{Box 1 } & \textbf{Box 2 } & \textbf{Box 2} \\
  \hline
      kpc & ($\Delta V < 500$) km s$^{-1}$ & ($500 < \Delta V < 1000$) km s$^{-1}$ & ($\Delta V < 500$) km s$^{-1}$ & ($500 < \Delta V < 1000$) km s$^{-1}$ \\
      \hline
    \hline
    $ < 20$ & 355 & 21 & 403 & 72 \\ 
    $20 < r_p < 50$ & 719 & 113 & 1080 & 318 \\
    $50 < r_p < 100$ & 1181 & 204 & 1967 & 625 \\
    $100 < r_p < 250$ & 2548 & 558 & 6072 & 1935 \\ 
   $250 < r_p < 500$ & 2442 & 558 & 7969 & 2914\\
    $500 < r_p < 1000$ & 2685 & 738 & 11411 & 4991\\
    \hline
   Tot. number of pairs & 9930 & 2192 & 28902 & 10855\\
    \hline
    \end{tabular}
\end{table*}

\subsection{Multiwavelength data}

We match the galaxies in pairs in Boxes 1 and 2 with other databases to study star formation and AGN activity, as shown in Table \ref{tab:survey}. The cross-match radius selected gives the best compromise between the number of lost counterparts and the number of spurious matches after testing for various radii (1" - 15") the number of counterparts using the ’best’ and ’all’ matches. In order to measure $(NUV-r)_0$ colour as a proxy of star formation, we extract NUV magnitudes from GALEX DR7 All Sky Imaging Survey (AIS) \citep{bianchi2017revised} using a $5''$ cross-matching radius. This is similar to the $4.5''$ size of the GALEX point spread function in NUV: 86\% of the GALEX sample falls within a $3”$ search radius. 

Optical photometry for our sources comes from the Pan-STARRS survey \citep{Chambers2016,Flewelling2016} and from the SkyMapper Southern Sky Survey DR4 \citep{Keller2007,Onken2024}
for galaxies below $\delta=-30^{\circ}$. We use a cross-match radius of $2''$ in optical, similar to the stated resolution of unWISE images \citep{Lang2014}.

To search for AGN in our galaxies, we cross-matched our sample with X-ray and radio all-sky surveys, using the eROSITA  \citep{Sunyaev2021} and NVSS, \citep{Condon1998}, respectively. To cross-match with eROSITA data, we used a cross-match radius of $10''$ to measure X-ray fluxes in the two soft 0--2.3 keV and hard 2.0--5.0 keV bands. The Half Energy Width of the eROSITA Point Spread Function (PSF) as reported by \cite{merloni2024srg} is $\sim 30''$: 58\% of the eROSITA sources lie within a search radius of $5”$. For NVSS data, we used a cross-match radius of $10''$ to measure radio fluxes at 21\,cm (1.4\,GHz). The NVSS beam is $45''$: 57\% of the NVSS sources are within $3”$ of the optical position. 

Foreground extinction in all bands was corrected using estimates from the NED. 

Finally, we obtain the emission line ratios and AGN classifications in optical for a subset of galaxies using theBaldwin, Phillips \& Terlevich (BPT) diagrams \citep{Baldwin1981}. We used the SDSS DR8 MPI-JHU catalogue\footnote{https://www.sdss4.org/dr17/spectro/galaxy\_mpajhu/} \citep{brinchmann2004physical} and a cross-match radius of $2''$. The total number of counterparts found in each survey for each box is shown in Table \ref{tab:survey}. 

\begin{table*}[]
    \centering
 \caption{Number and fraction of galaxies in pairs detected in  GalexPanSky (combination of GALEX, Pan-STARRS and SkyMapper), eROSITA, NVSS, and SDSS DR8 MPA-JHU}.
    \label{tab:survey}
    \begin{tabular}{lccccccc}
    \hline
       \textbf{Survey (Ref.*)} &\textbf{Data}& \textbf{Band}&\textbf{Radius (arcsec)} & \textbf{Box 1**} & \textbf{Box 2**}\\
        \hline
        GalexPanSky (1)& UV & - &5 &16158 (67\%) & 44505 (57\%)\\
         eROSITA (2) & X-ray & soft (0.2-2.3 keV) &10 & 719 (3\%) & 1876 (2\%)\\
         eROSITA (2)& X-ray & hard (2.3-5 keV) & 10 &  400 (1.66\%) & 1065 (1.34\%) \\
         NVSS (3)& Radio& 21\,cm (1.4\,GHz) & 5 & 2522 (10\%) & 3582 (5\%) \\
         SDSS DR8 MPA-JHU (4) & Optical& - &2& 9643 (40\%) & 38215 (48\%) \\
       
    \hline
    \end
    {tabular}\\
   \small{*Refers to: 1 - \citep{Chambers2016,bianchi2017revised,Onken2024}, 2 -  \cite{merloni2024srg}, 3 - \cite{Condon1998},\\ and 4 - \citep{brinchmann2004physical}. **Fractions are measured by taking into account the total number of galaxies in pairs in Boxes 1 and 2}. 
\end{table*}

\section{Results}

\subsection{Star formation}
\label{sec_SF}

\begin{figure}
  \centering
   \includegraphics[width=1.0\linewidth]{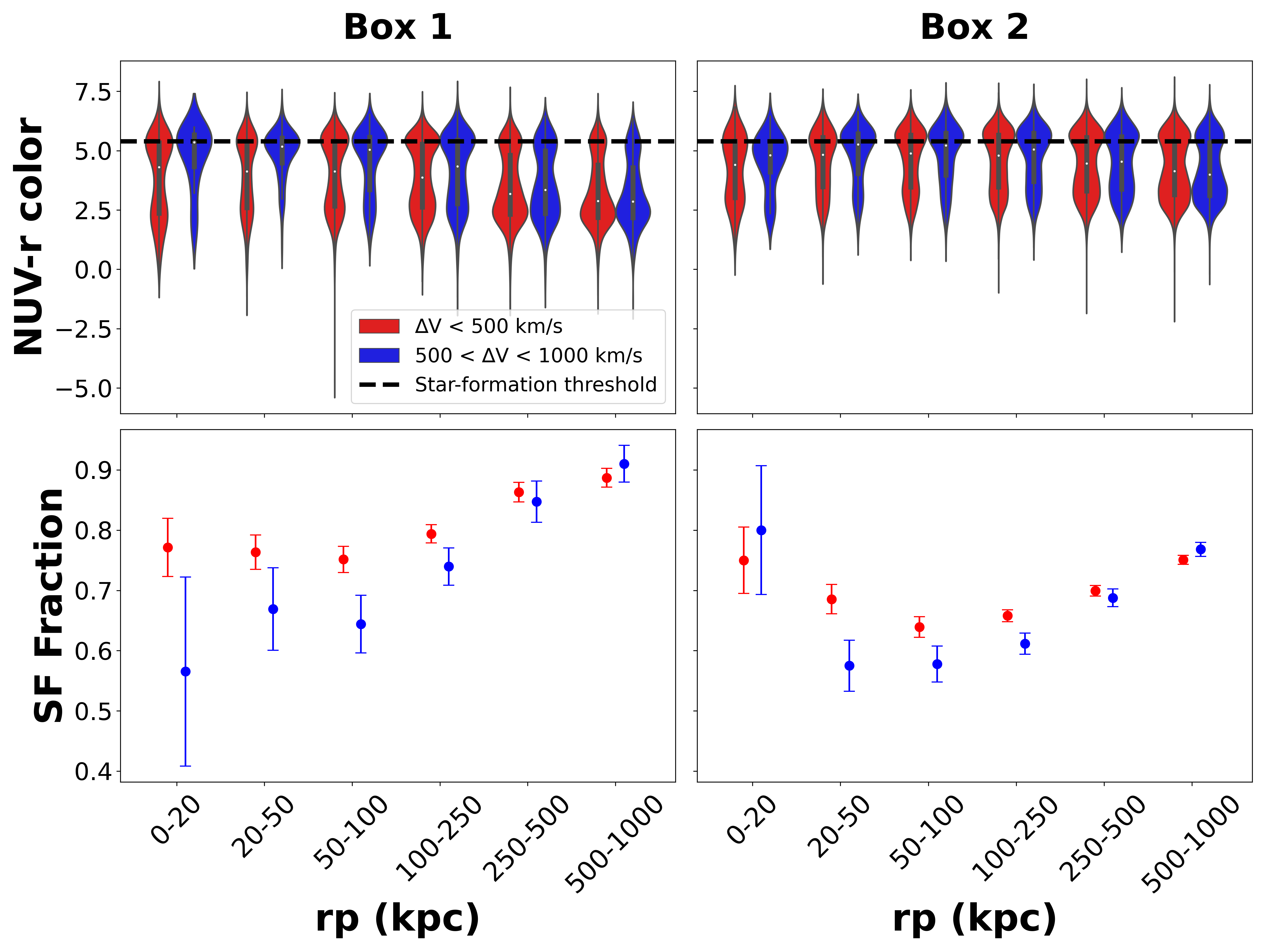}
    \caption{Top: Violin plots of the distribution of 
    $(NUV-r)_0$ colours for galaxies in the lower mass
    sample (Box 1: $-21 < M_{W1} < -24$ and \(z < 0.03\)) on the left and higher mass sample (Box 2: $-24 < M_{W1} < -27$ and z \(<\) 0.10) on the right as a function of pair separation (on the X-axis) and velocity difference (red for $< 500$ km s$^{-1}$ and blue for $500 < \Delta V < 1000$ km s$^{-1}$). Note that we have slightly offset the two velocity samples for clarity. Galaxies below the dashed line are classified as star-forming. Bottom: The mean fraction of star-forming galaxies as a function of pair separation and velocity difference (same colour scheme as top panels) in Box 1 (left) and Box 2 (right). For the error bars see the text.}
    \label{violin1}
\end{figure}

We use the $(NUV-r)_0$ colour as a proxy for the degree of star formation present in each pair member. Following \cite{Yi2008}
and \cite{Hernandez2013} we use $(NUV-r)_0 = 5.4$ as the discriminant between star-forming ($< 5.4$) and quiescent ($> 5.4$) galaxies. In Table \ref{table:SF&active}, we show the total number of galaxy pairs that have either one or both pair members star-forming. In Figure~\ref{violin1} (top) we plot violin plots \citep{Hintze1998} to display the probability distribution of pair galaxies as a function of $(NUV-r)_0$ colour for each separation and velocity difference. Violin plots visualise the probability distribution of numerical data. The width of the violin represents the density of data points at that specific value. The white dot inside the violin indicates the median while the thick dark central bar is the interquartile range (IQR). At the ends whiskers extend from the IQR to the largest and smallest values within $1.5\times$IQR beyond which they are considered outliers. The overall shape of the violin reveals the distribution's form, including peaks, flatness, or skewness. 
The central black bar allows us to observe the data's spread and dispersion.  We then plot the mean fractions of star-forming galaxies as a function of separation $r_p$ and velocity difference in Figure~\ref{violin1} (bottom). The errors in the fractions were measured following Poisson’s distribution as \(\sigma = \sqrt{n}/N\), where n is the fraction of galaxies deemed to be star-forming (i.e., $(NUV-r)_0 < 5.4$) and N is the total number of galaxies per bin. This method was applied for bins where N is greater than 30. In cases where N is less than 30, we apply the distribution described in equation A1 of \cite{Burgasser2003}. In this approach, we estimate the upper bound as 84th percentile, and the lower bound as the 16th percentile. The corresponding fractions with their errors are represented in Table \ref{table:SFR}.

For the lower-mass sample (Box 1), we notice, for both velocity ranges, that the fraction of star-forming galaxies
declines as the projected separation decreases, flattening or slightly rising in the closest separation regime we 
consider ($r_p$ $< 20$ kpc). We observe a similar behaviour for the higher mass sample (Box 2), but the increase in star formation at closer
separations ($r_p$ $< 20$ kpc) is more pronounced.

\begin{figure}
    \centering    \includegraphics[width=1\linewidth]{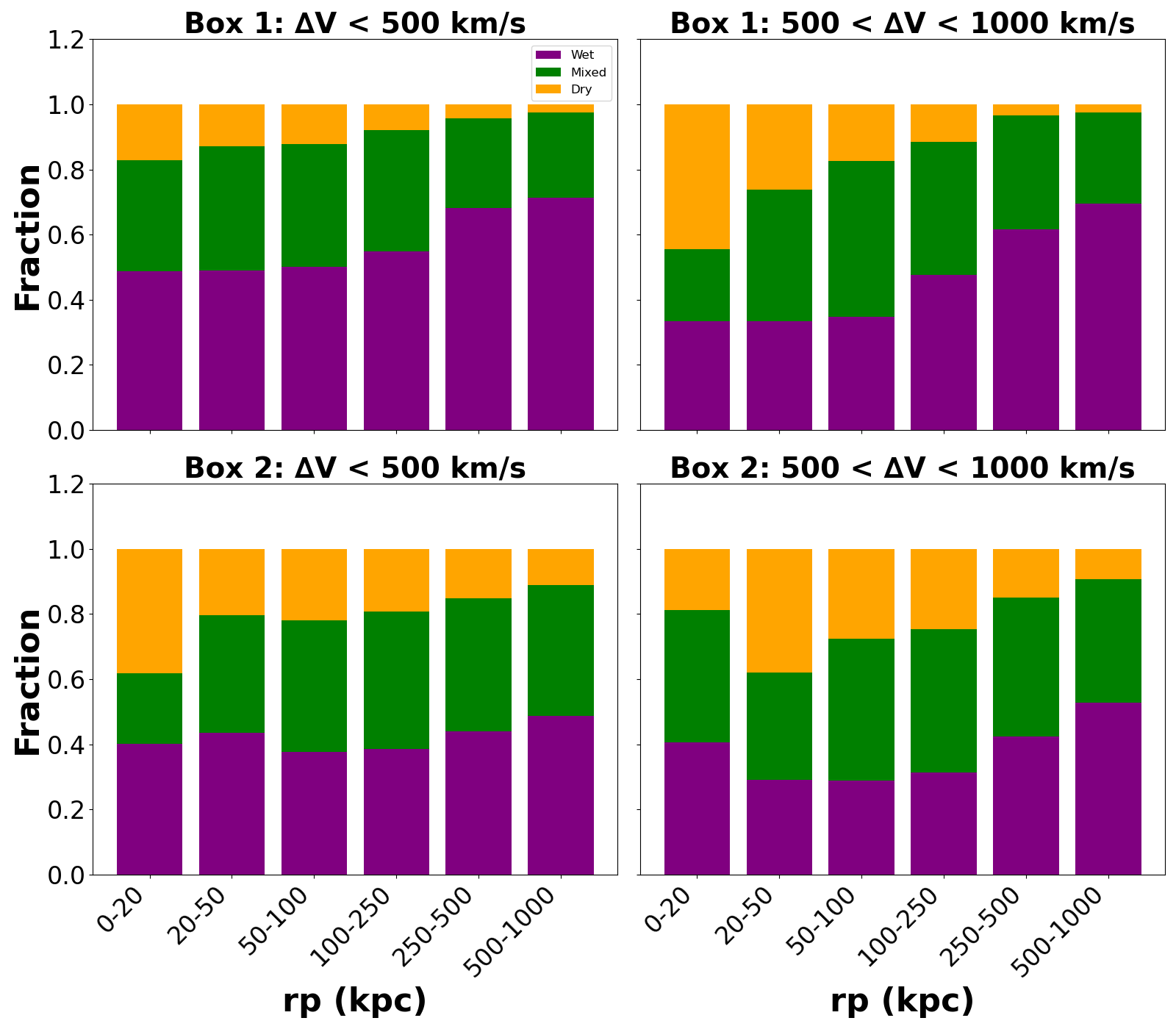}
    \caption{Cumulative histogram of the wet, mixed, and dry merger fraction as a function of projected separation and velocity difference.}
    \label{dry}
\end{figure}

\begin{figure}
    \centering    \includegraphics[width=1\linewidth]{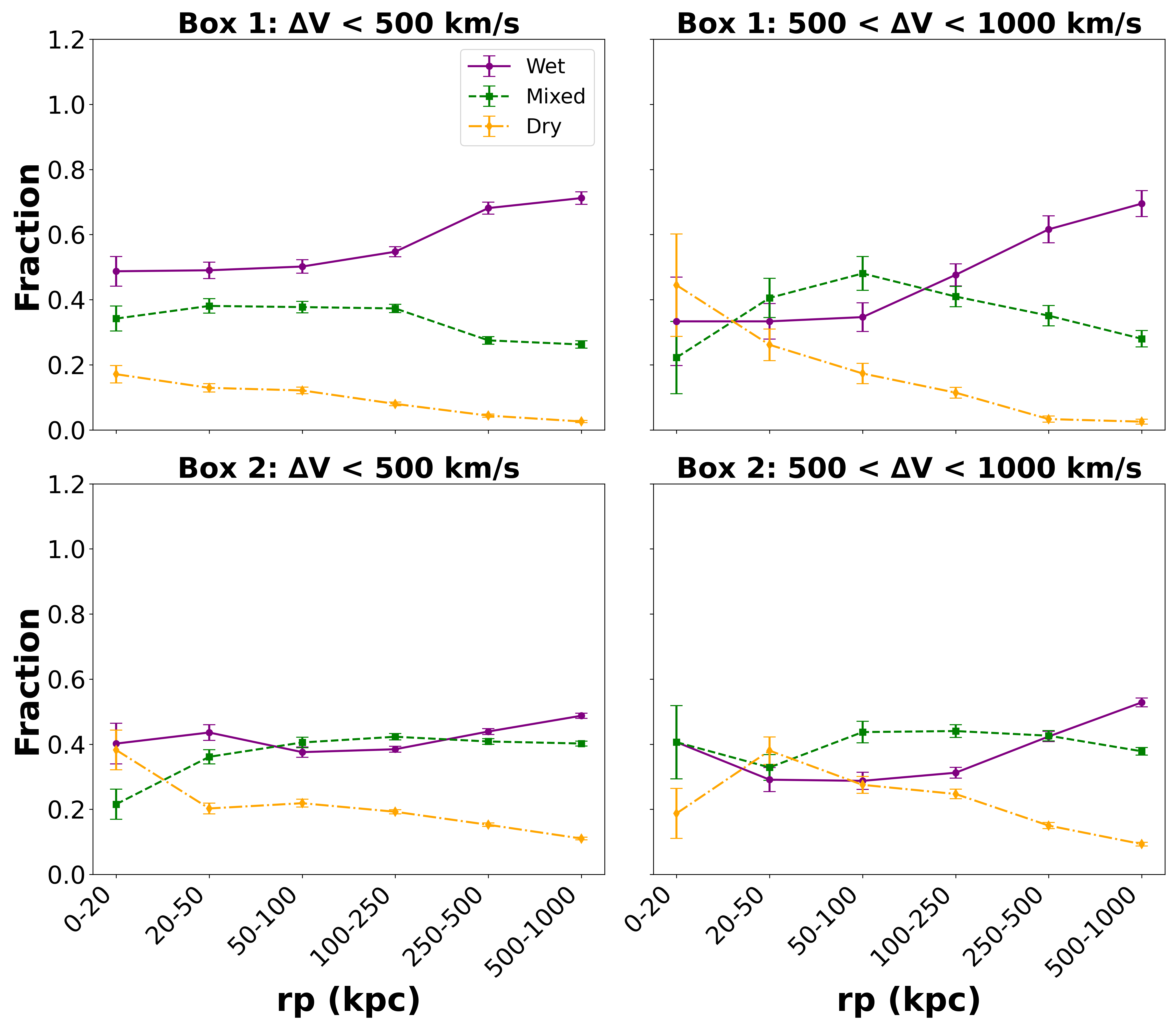}
    \caption{Mean fractions of the wet, mixed, and dry merger as a function of projected separation and velocity difference.}
    \label{wet}
\end{figure}

For the closer separations, the increase in star formation
may be explained by shocks and tidal stresses compressing the interstellar medium. However, at larger separation,s one observes an apparent quenching of star formation from the values in the more distant galaxies in pairs. While interactions can suppress star formation, at $r_p > 50$ kpc we see no evidence that galaxies in pairs are strongly interacting. One possibility could be that the apparent quenching reflects the somewhat denser environments in which closer pairs may reside.

In Fig.~\ref{dry} we show the fractions of pairs classified as wet, mixed, and dry mergers as a function of projected separation and velocity difference. We classify pairs as wet mergers when both pair members have $(NUV-r)_0 < 5.4$ (star-forming), as mixed mergers when one of the pair members is star forming, and as dry mergers when none of the pair members is star forming. In agreement with Fig.~\ref{violin1}, we see a slight increase in the fraction of dry mergers and a slight decrease in that of wet mergers as $r_p$ decreases for both $\Delta V$. In all cases, there is a slight increase in dry pairs (between red galaxies) as a function of decreasing separation and a corresponding decrease in the fraction of wet pairs, while mixed pairs are not strongly affected.

\begin{table*}[]
   \centering
       \caption{Total number of galaxy pairs that have one or both galaxy pair members star-forming, and active in X-rays, radio, and infrared for Boxes 1 and 2.}
    \label{table:SF&active}
 \begin{tabular}{l|ccc|ccc}
\hline
 \textbf{Data} & \multicolumn{3}{c}{\textbf{Box 1}} & \multicolumn{3}{c}{\textbf{Box 2}} \\
 \hline
& \textbf{Tot. N } & \textbf{Both} & \textbf{One } & \textbf{Tot. N} & \textbf{Both} & \textbf{One} \\
&\textbf{of pairs} &\textbf{members }& \textbf{pair member}& \textbf{of pairs} & \textbf{members} &  \textbf{pair member}\\
\hline
\hline
Star-forming galaxy & 10392 & 2949  & 7443  & 25711  & 5742 & 19969 \\
X-ray (soft) AGN & 22 & 0 & 22 & 792 & 12 & 780 \\
X-ray (hard) AGN & 29 & 0 & 29 & 582 & 5 & 577 \\
Radio AGN & 27 & 0 & 27 & 1408 & 11 & 1397 \\
Infrared AGN & 55 & 0 & 55 & 377 & 4 & 373 \\
\hline
\end{tabular}
\end{table*}

\subsection{Active Galactic Nuclei}
\label{sec_AGN}

\begin{figure}[h!]
    \centering
    \includegraphics[width=1.0\linewidth]{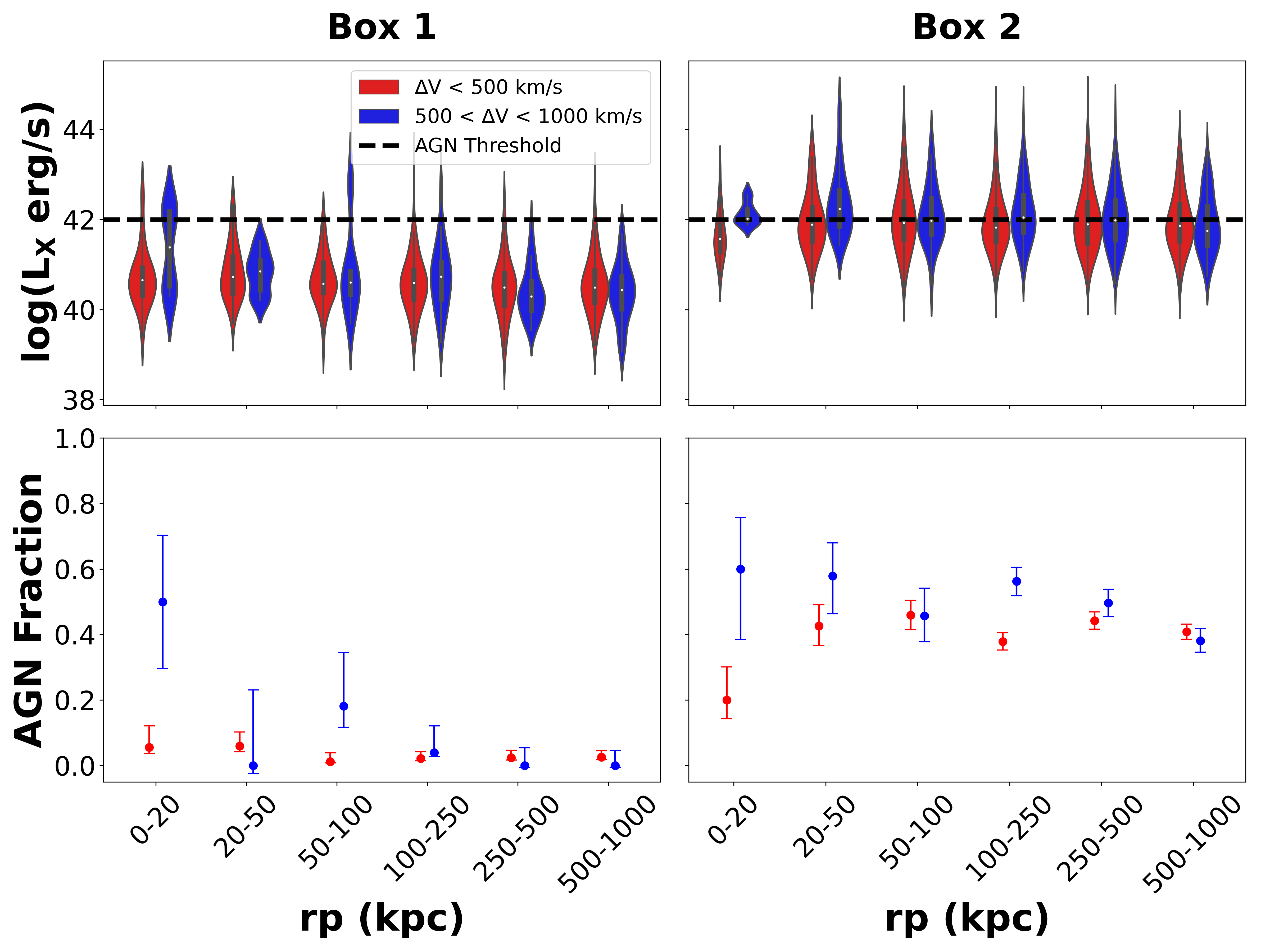}
    \caption[Soft band (0.2-2.3 keV), X-ray luminosities]{ Distribution of soft (0.2-2.3 keV) X-ray luminosities (top panels) with corresponding X-ray detected mean AGN fractions (bottom panels). Galaxies above the dashed line are classified as AGN. The layout of this figure is as in Fig.~\ref{violin1}.}
    \label{violinX1}
\end{figure}

\begin{figure}[h!]
    \centering
    \includegraphics[width=1.0\linewidth]{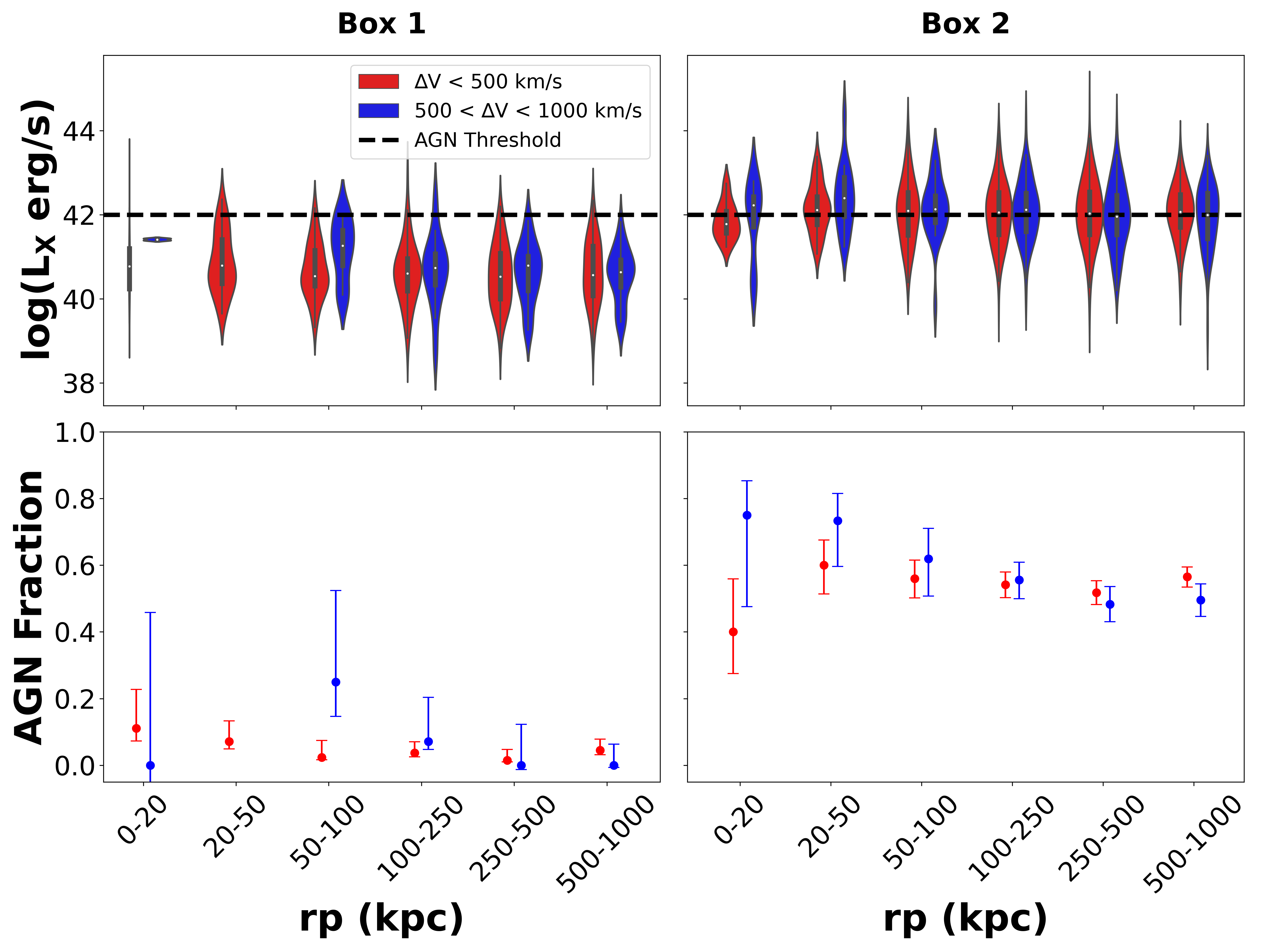}
    \caption{Same as Fig.~\ref{violinX1}, but in hard (2-5 keV) X-ray band.}
    \label{violinX2}
\end{figure}

\begin{figure}[h!]
    \centering
    \includegraphics[width=1.0\linewidth]{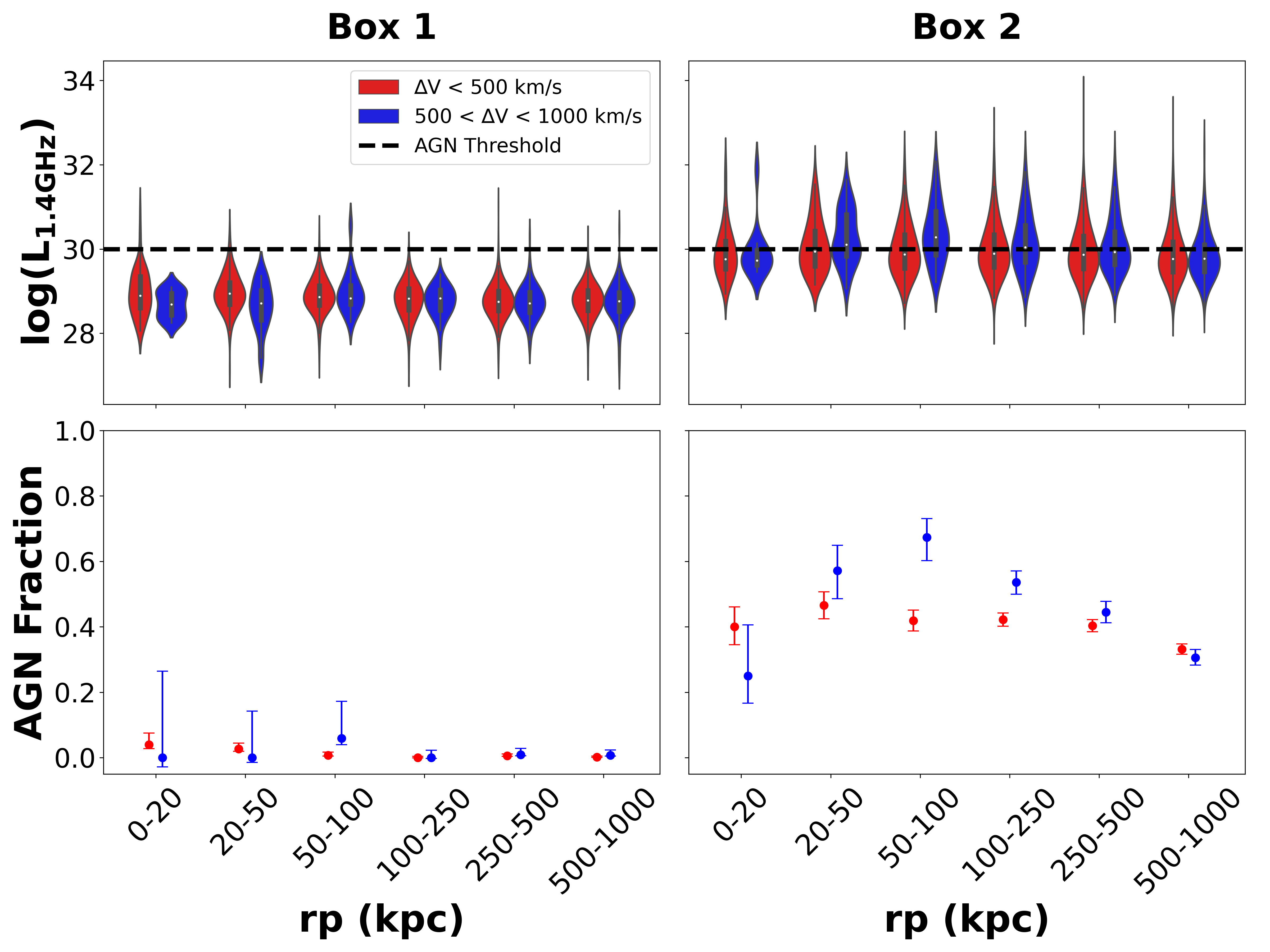}
    \caption{Distribution of radio luminosity (top) and the mean fraction of radio-selected AGN (bottom). Galaxies above the dashed line are classified as AGN. The layout of this figure is as in Fig.~\ref{violin1}.}
    \label{violinradio}
\end{figure}

\begin{figure}[h!]
    \centering
    \includegraphics[width=1.0\linewidth]{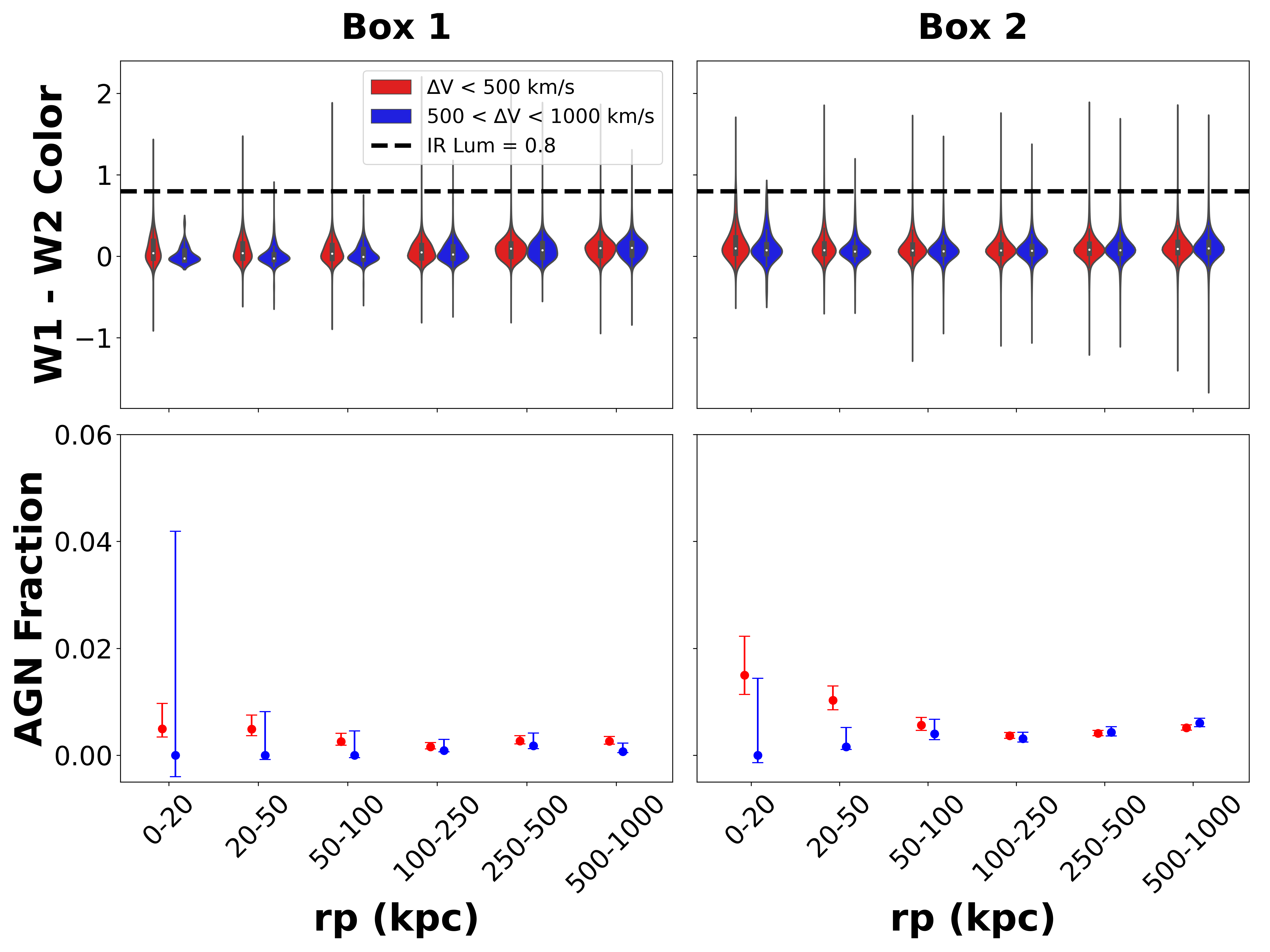}
    \caption{Distribution of $W1-W2$ colours (top) for our pair samples and derived mean AGN fractions (bottom). Galaxies above the dashed line are classified as AGN. The layout of this figure is as in Fig.~\ref{violin1}.}
    \label{violinIR}
\end{figure}

We select AGN in each pair member using three different methods:
\begin{itemize}
\item X-ray detections in the eROSITA soft (0--2.3 KeV) and hard (2--5 KeV) bands, where we adopt a luminosity threshold of $L_X > 10^{42}$ ergs s$^{-1}$ as our definition of a galaxy hosting an AGN for both soft and hard bands \citep{Brandt2015,Liu2022,Birchall2022}. We show the violin plots and the fractions of AGN as a function of separation for both mass ranges in Fig.~\ref{violinX1} for the soft band, and in Fig.~\ref{violinX2} for the hard band.
\item Radio data at 1.4GHz from the NVSS survey (only above $\delta=-40^{\circ}$), where we adopt a threshold of $L > 10^{30}$ ergs s$^{-1}$ Hz$^{-1}$ for detecting AGN \citep{Sadler2002, Best2005,Mauch2007,Condon2019}. The violin plots and the fractions of radio-detected AGN are shown in Fig.~\ref{violinradio}.
\item WISE data, where we use a color $W1-W2 > 0.8$ criterion to identify AGN \citep{Stern2012,Assef2013,Assef2018}. The distribution of colours as a violin plot and the fractions of IR-detected AGN are shown in Fig.~\ref{violinIR}.
\end{itemize}
Table \ref{table:SF&active} shows the total number of galaxy pairs that have either one or both pair members as AGN selected in X-rays (soft and hard), radio, and infrared. We represent the fractions and their errors of AGN as a function of projected separation and velocity difference in Tables \ref{table:x_ray_soft}, \ref{table:x_ray_hard}, \ref{table:radio}, and \ref{table:ir}.

As expected, in all cases, AGN fractions are higher in the more massive sample (Box 2; e.g., \citealt{Kauffmann2003,Vitale2013,Mahoro2017}). Other works have reported similar trends: for instance, \cite{Pimbblet2012} finds an increase in AGN fraction by a factor of about 3 across the mass range we consider here. However, in all the AGN detection methods used,
we observe only a weak trend of increased AGN activity with decrease in projected separation: the significance of the increase in AGN fraction towards lower $r_p$ is weak. There are some marginal trends for an increase in AGN fraction at closer separations for objects in our highest velocity separation bins, but we do not consider these as providing strong evidence in either case. This is consistent
with recent works that favour secular processes as the trigger for most local AGN, rather than major mergers, except possibly for the more luminous quasars \citep{Ho2009,Hopkins2010,Cisternas2011,Menci2014,Villforth2017,Ellison2019,Zhao2022}.

\subsection{Star formation and AGN by emission line indices}

\begin{figure*}
\sidecaption 
    \includegraphics[width=12cm]{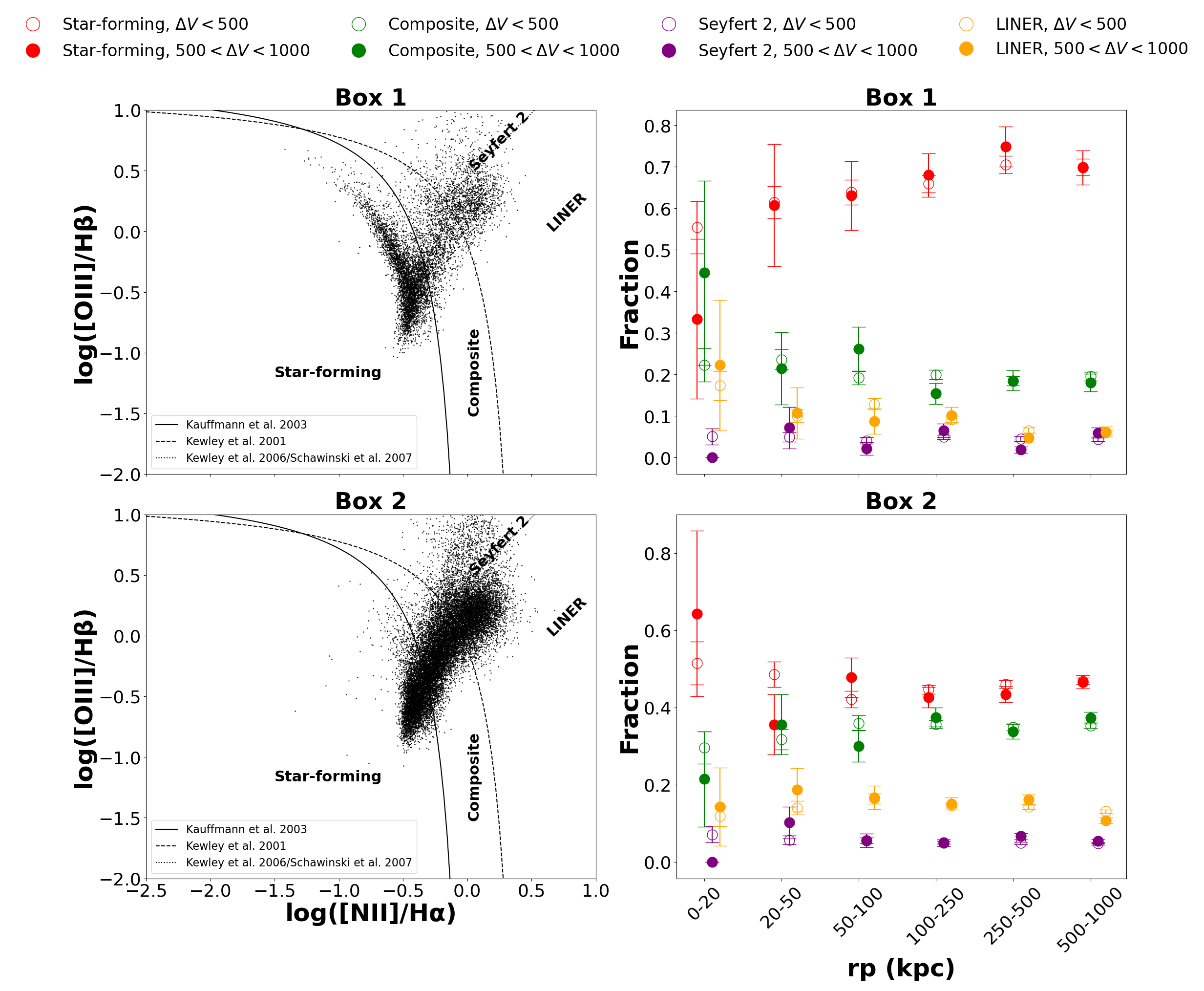}
\caption{Left: The BPT-NII diagrams for our pair samples in Box 1 (top) and Box 2 (bottom), for all analysed separations and velocity differences. Right: Fraction of galaxies classified as star-forming, composite, Seyfert 2 and LINER as a function of pair separation and velocity difference in Box 1 (top) and Box 2 (bottom). See legend in figure to identify the samples.}
    \label{BPT1}
\end{figure*}

With the emission line indices from SDSS DR8 MPA-JHU, \citep{York2000,Abdorrouf2022} we identify star-forming and active galaxies using the BPT-NII diagram \citep{Baldwin1981}. These line strengths have been derived by \cite{Kewley2001}, \cite{Kauffmann2003}, \cite{Kewley2006}, and \cite{Schawinski2007} for SDSS galaxies. Figure~\ref{BPT1} shows the diagram (left panels) for
all our pair members including all separations and velocity differences, and the fraction (right panels) of star-forming, composite, Seyfert 2, and Low-Ionisation Nuclear Emission-line Region (LINER) galaxies as a function of projected separation and velocity difference for both mass ranges we consider. In general, we find very similar trends as in sections~\ref{sec_SF} and \ref{sec_AGN}. The closest pair members with $r_p < $\,20\,kpc show a modest increase in the fraction of star-forming galaxies against the next separation, with a modest increase towards the larger separations, as observed in sec.~\ref{sec_SF}. Similarly, we find very weak trends in Seyfert 2 and LINER fractions with pair members separation, in agreement with the previous indicators used above.

\begin{figure*}
\sidecaption 
\includegraphics[width=12cm]{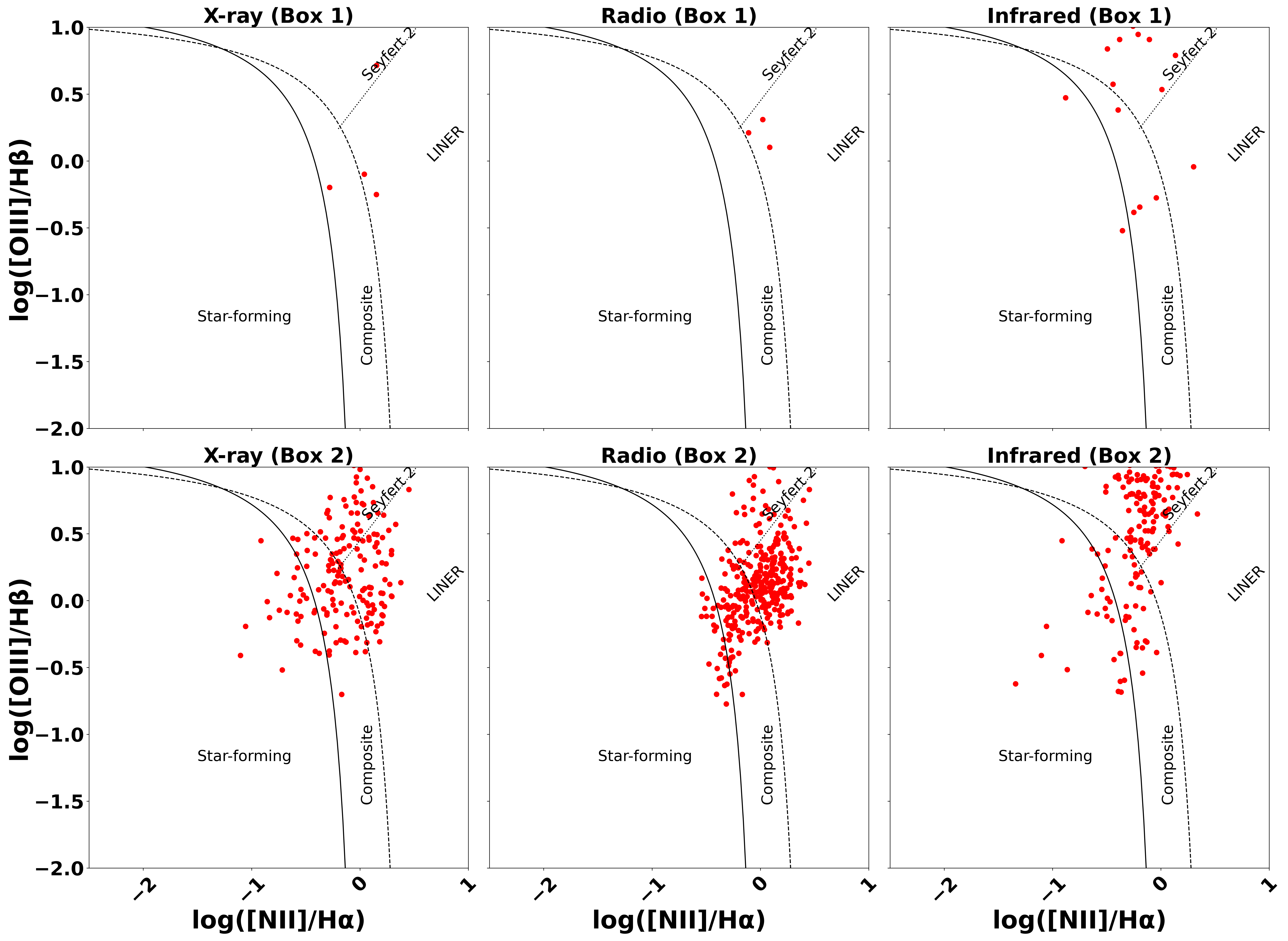}
    \caption{Comparison of BPT-NII diagram to our AGN samples detected in soft (0.2-2.3 keV) X-rays (left), radio (middle) and infrared (right) in Box 1 (top) and Box 2 (bottom).}
    \label{BPT2}
\end{figure*}

We compare the three AGN-detection methods using BPT diagrams in Fig.~\ref{BPT2}. There are only a few objects in the lower mass sample (Box 1). In the higher mass sample (Box 2), X-ray AGN selection seems to be the least biased approach detecting both Seyfert 2 and LINERs, while radio selection tends to
favour LINERs, and infrared selection prefers Seyfert 2 galaxies, which is in line with previous studies \citep[e.g.,][]{Koutulidis2013,Ji2022}.

\begin{table*}[!hbtp]
   \centering
       \caption{Fraction of star-forming galaxies in pairs per separation and velocity bin}.
    \label{table:SFR}
    \begin{tabular}{lcccc}
  \hline
     Separation range ($r_p$) & Fraction in Box 1 & Fraction in Box 1 & Fraction in Box 2  & Fraction in Box 2 \\
      kpc & $\Delta V < 500$ km s$^{-1}$ & $500 < \Delta V < 1000$ km s$^{-1}$ & $\Delta V < 500$ km s$^{-1}$ & $500 < \Delta V < 1000$ km s$^{-1}$ \\
      \hline
    \hline
    $ < 20$ &  $0.771 \pm 0.048$ &$0.565 \pm 0.16$ & $0.750 \pm 0.055$ & $0.800 \pm 0.107$ \\ 
    $20 < r_p < 50$ &  $0.764 \pm 0.028$ &$0.669 \pm 0.069$ & $0.685 \pm 0.024$ & $0.575 \pm 0.042$ \\
    $50 < r_p < 100$ &  $0.751 \pm 0.022$ &$0.644 \pm 0.048$ & $0.639 \pm 0.017$ & $0.578 \pm 0.030$ \\
    $100 < r_p < 250$ &  $0.794 \pm 0.015$ &$0.740 \pm 0.031$ & $0.658 \pm 0.010$ & $0.612\pm 0.017$ \\ 
   $250 < r_p < 500$ &  $0.887 \pm 0.016$ &$0.0.847 \pm 0.034$ & $0.699 \pm 0.009$ & $0.688\pm 0.015$\\
    $500 < r_p < 1000$ &  $0.887 \pm 0.016$ &$0.910 \pm 0.030$ & $0.751 \pm 0.008$ & $0.768\pm 0.012$\\
    \hline
    \end{tabular}
\end{table*}

\subsection{Morphologies of pair galaxies}
\label{morphology}
Close galaxy pairs are only one of the methods employed to find possible merging galaxies. The structure and morphology of galaxies can be easily altered by close encounters, leaving asymmetric features such as tidal tails \citep{Toomre1972,Huang2022}. Morphological disturbances
are common in merging and post-merging systems (e.g., \citealt{Conselice2003,Pawlik2016}). As mentioned above, \cite{DePropris2007} looked for signs
of morphological disturbance in dynamically close pairs and estimate a lower limit of 35\% to the true merger fraction, while 53\% of the close pairs in \cite{Desmons2023} exhibit tidal features. It was suggested that samples that are restricted to interacting
and merging galaxies show a greater prevalence of starbursts and AGN (e.g., \citealt{Cezar2024,Robin2024}).

\begin{figure}
    \centering
    \includegraphics[width=1\linewidth]{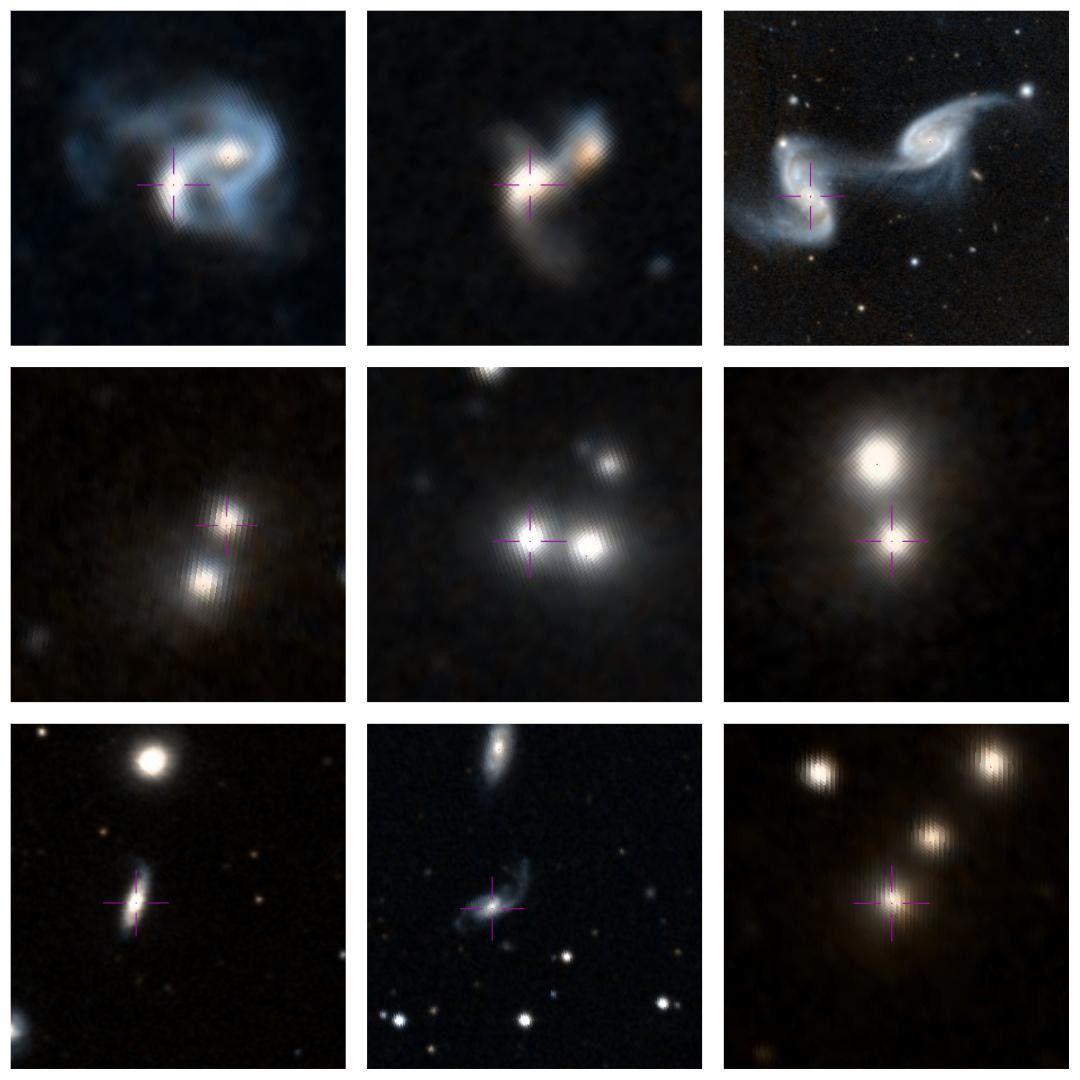}
    \caption{Examples of strongly interacting (top), weakly interacting (middle) and undisturbed galaxies in pairs (bottom) from PanStarrs images of our sample.}
    \label{images}
\end{figure}

\begin{figure}
    \centering    \includegraphics[width=1\linewidth]{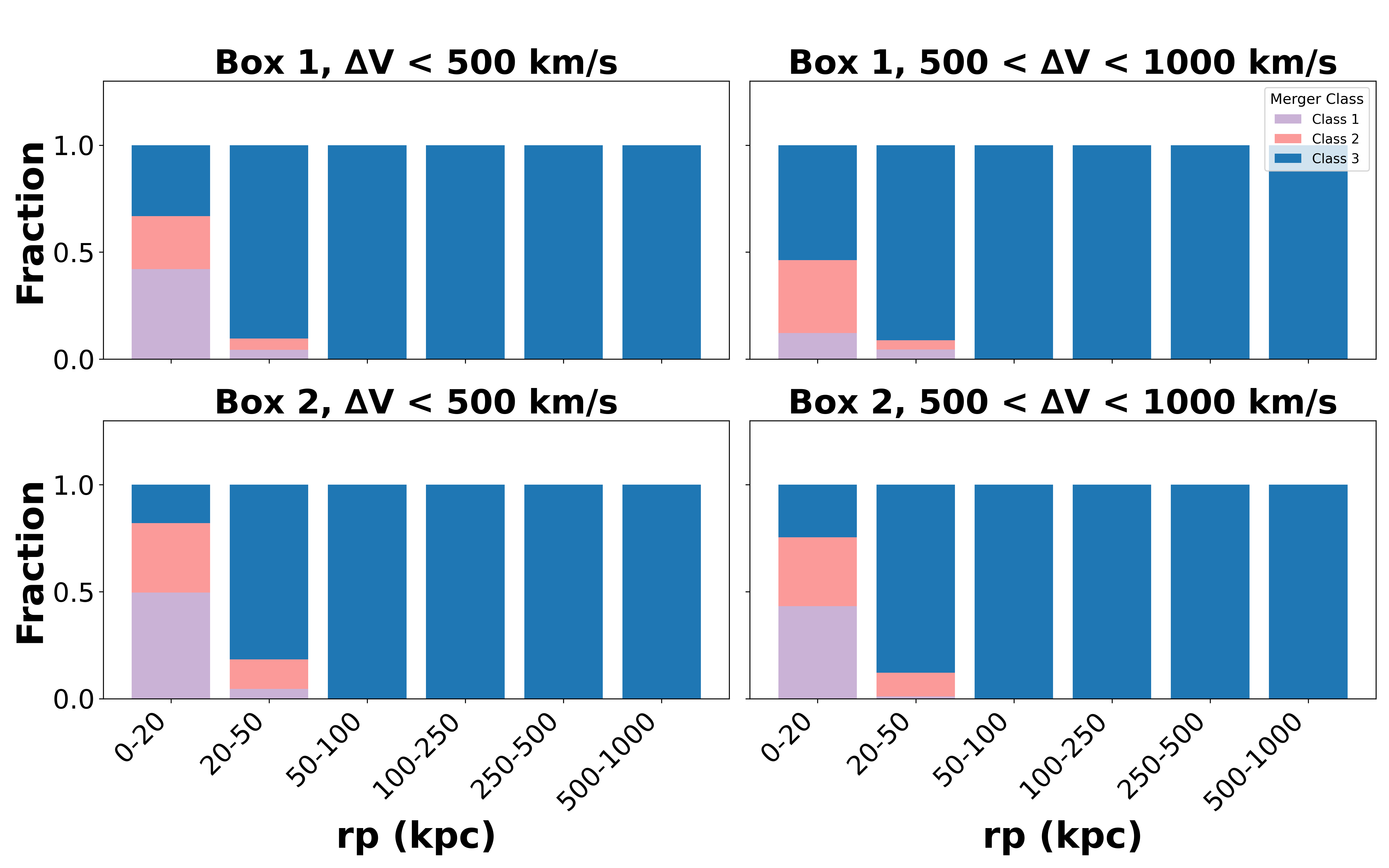}
    \caption{Fraction of galaxies corresponding to the classes 1, 2, and 3 for strongly
interacting, weakly interacting, and not interacting, respectively, for boxes 1 (top) and 2 (bottom) for two velocity differences (left and right plots).}
    \label{fig:merge}
\end{figure}

\begin{figure}
    \centering
    \includegraphics[width=1\linewidth]{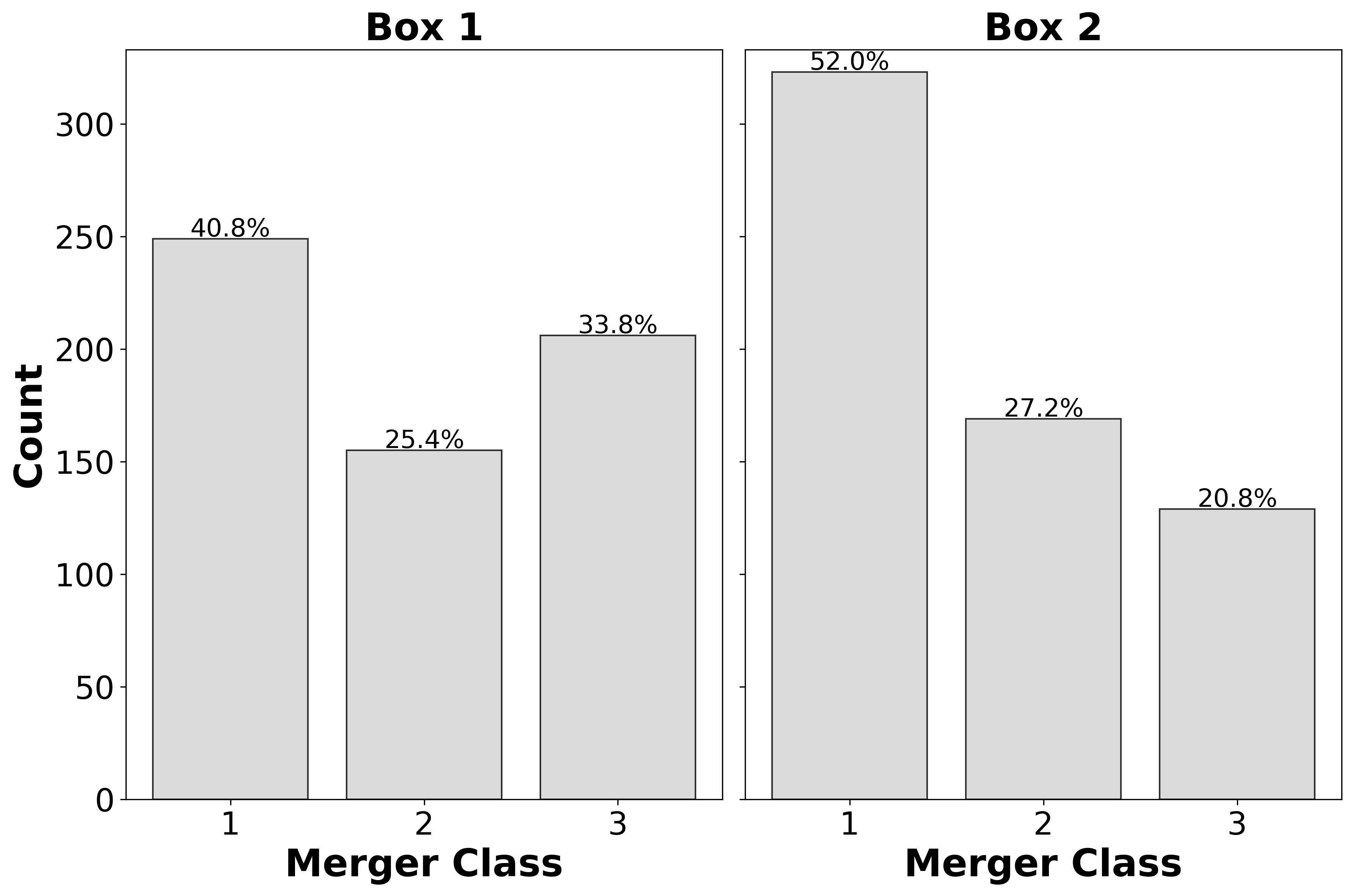}
    \caption{Merger class distribution for box 1 (left), and box 2 (right). Numbers
1, 2, and 3 correspond to strongly interacting, weakly interacting, and non-interacting galaxies
pairs, respectively, in the closest separations ($rp < 20 kp$ and $\Delta$ V \(<\) 500 km/s)}
    \label{fig:close_merger}
\end{figure}

We classified galaxies in pairs using images from the Pan-STARRS and Skymapper surveys. We classified galaxy pair members into three classes, as strongly interacting, weakly interacting, and non-interacting (or undisturbed), as shown in Fig.~\ref{images}. Strongly interacting galaxies as in the top row of Fig.~\ref{images} are observed to have visible morphological disturbances, irregular morphologies and prominent tidal tails. For weakly interacting galaxies, we only observe tidal tails and relatively small asymmetries in the light distribution, while undisturbed galaxies do not appear to show any distortion. We classified all objects in  Box 1, and 10\% of all objects were randomly selected in each pair separation class in Box 2. In Box 1, the total numbers of strongly interacting, weakly interacting, and non-interacting galaxy pairs are 167,125, and 11,830, respectively, while in Box 2 they are 233,304, and 5,010, respectively. Only pairs with $r_p < 50$ kpc are seen to contain disturbed systems as shown in Figure \ref{fig:merge} (although this may be affected by the limitation of the available imaging). Between 40\% and 50\% of pairs with  $r_p < 20$ kpc and $\Delta V < 500$ km/s are seen to be strongly interacting as shown in Figure \ref{fig:close_merger}, almost independently of the mass range selected, in agreement with \cite{Desmons2023}.

\begin{figure}
   \centering    \includegraphics[width=1\linewidth]{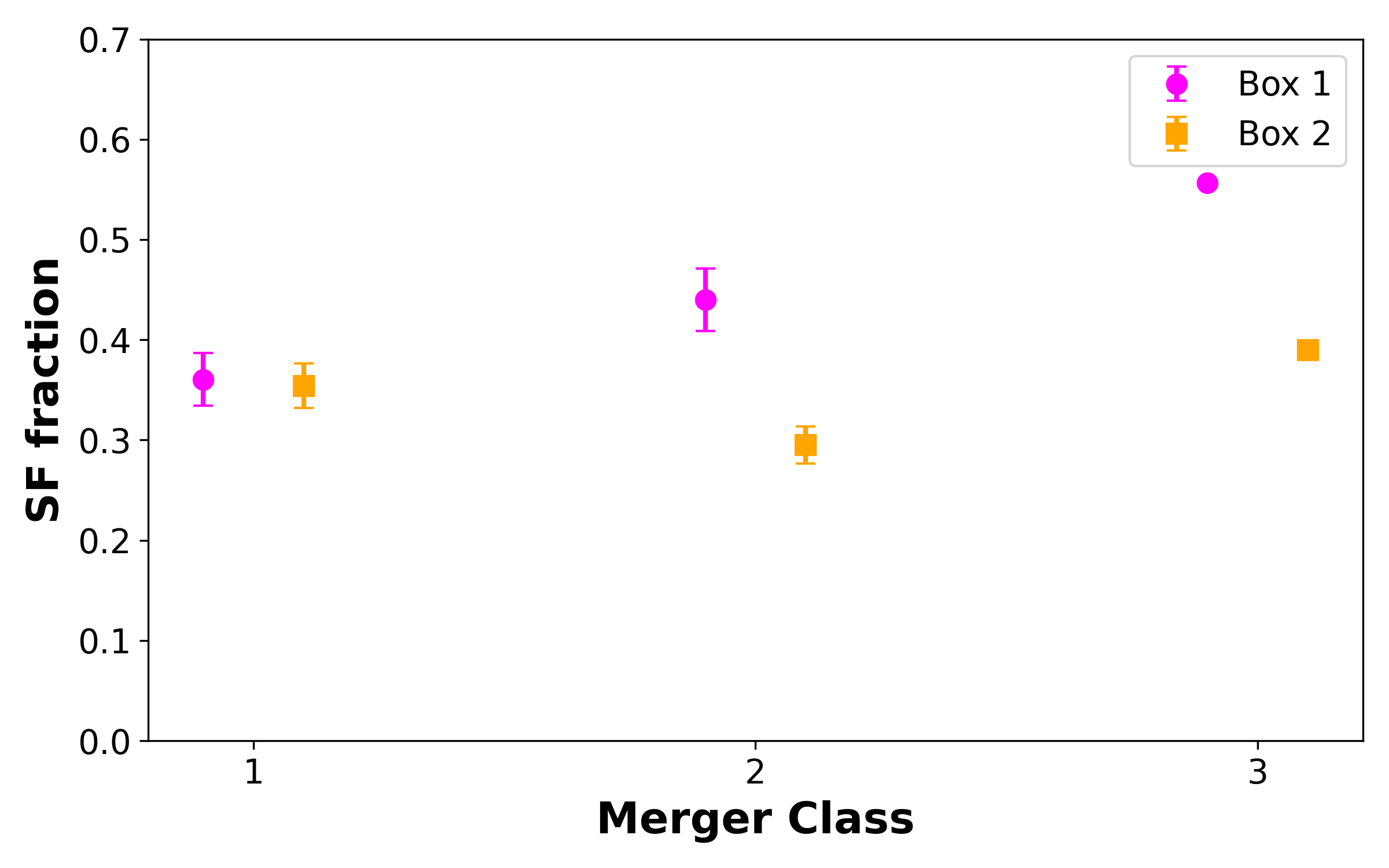}
   \caption{Mean fraction of star-forming galaxies in strongly interacting (class 1), weakly interacting (class 2) and undisturbed (class 3) galaxies in pairs in Box 1 (violet) and Box 2 (orange) at $r_p < 50$ kpc separation and all velocity differences. Errors are measured as in sec.~\ref{sec_SF}.}
   \label{realpairssfr}
    \end{figure}

We can now ask whether objects selected as interacting do show evidence of increased star formation activity or the presence of AGN, as in \cite{Ellison2024,Ellison2025}, for example.
We plot the mean fractions of star-forming galaxies using the $(NUV-r)_0 < 5.4$ diagnostic for all merger classes and for $r_p < 50$ kpc in Fig.~\ref{realpairssfr}. The trends we observe are similar for all three classes and to those we observe for the entire sample in Fig.~\ref{violin1}: i.e., there is no strong evidence that star formation increases in pair galaxies as they 'approach' each other, suggesting that even for closely separated pairs with clear signs of interaction the process does not suffice to induce strong star formation episodes.

\begin{figure}
   \centering
    \includegraphics[width=0.49\linewidth]{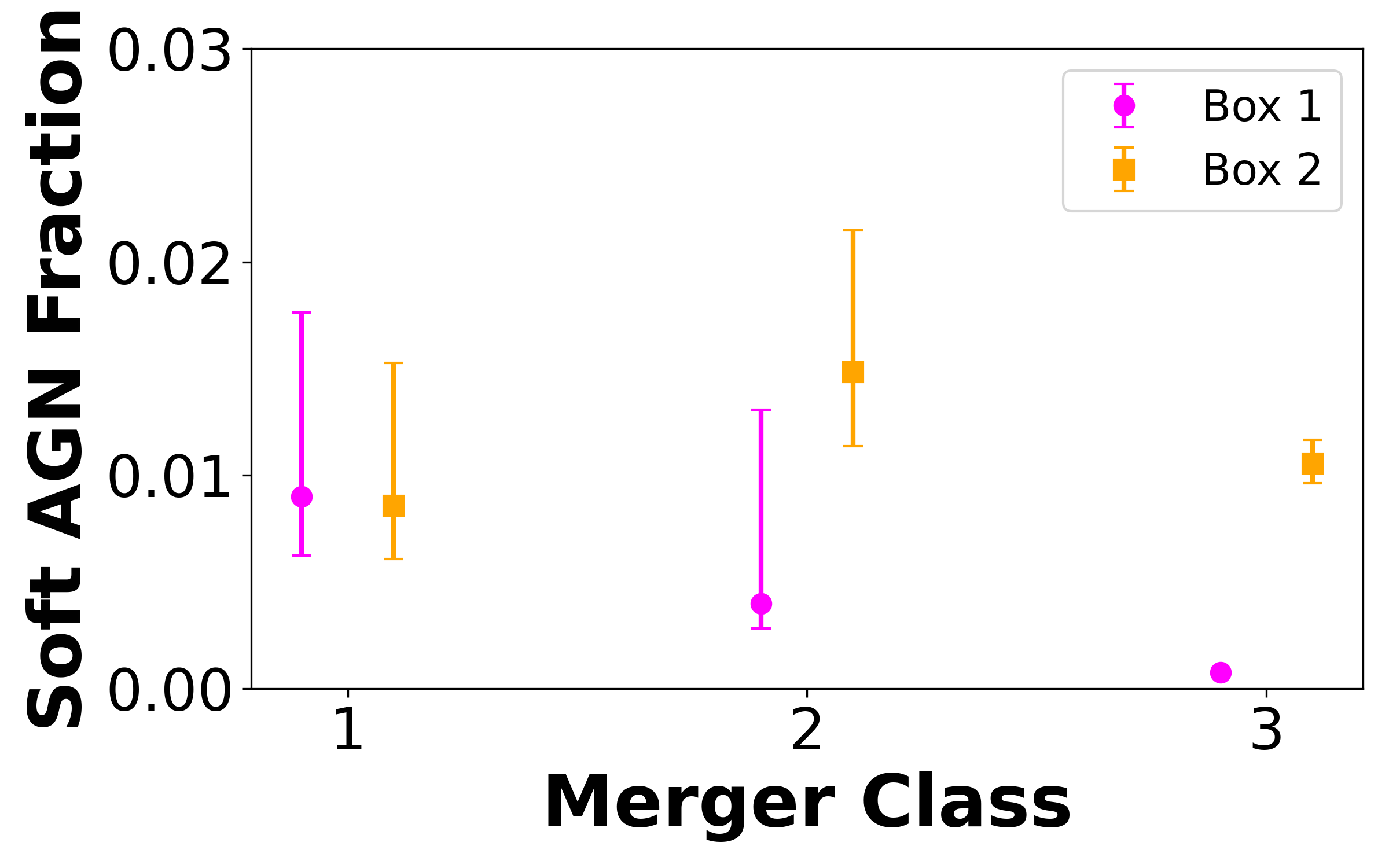}
    \includegraphics[width=0.49\linewidth]{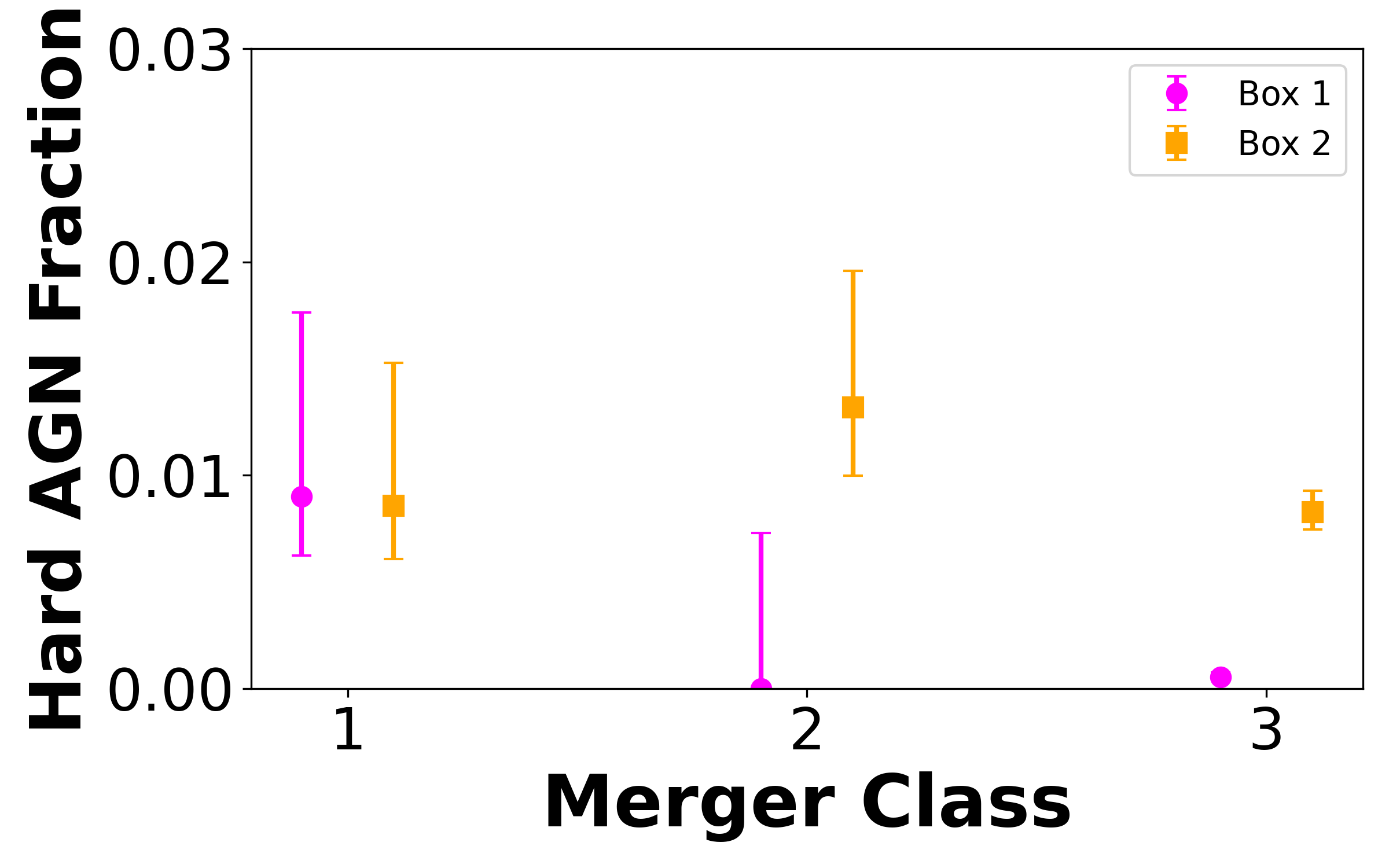}
    \includegraphics[width=0.49\linewidth]{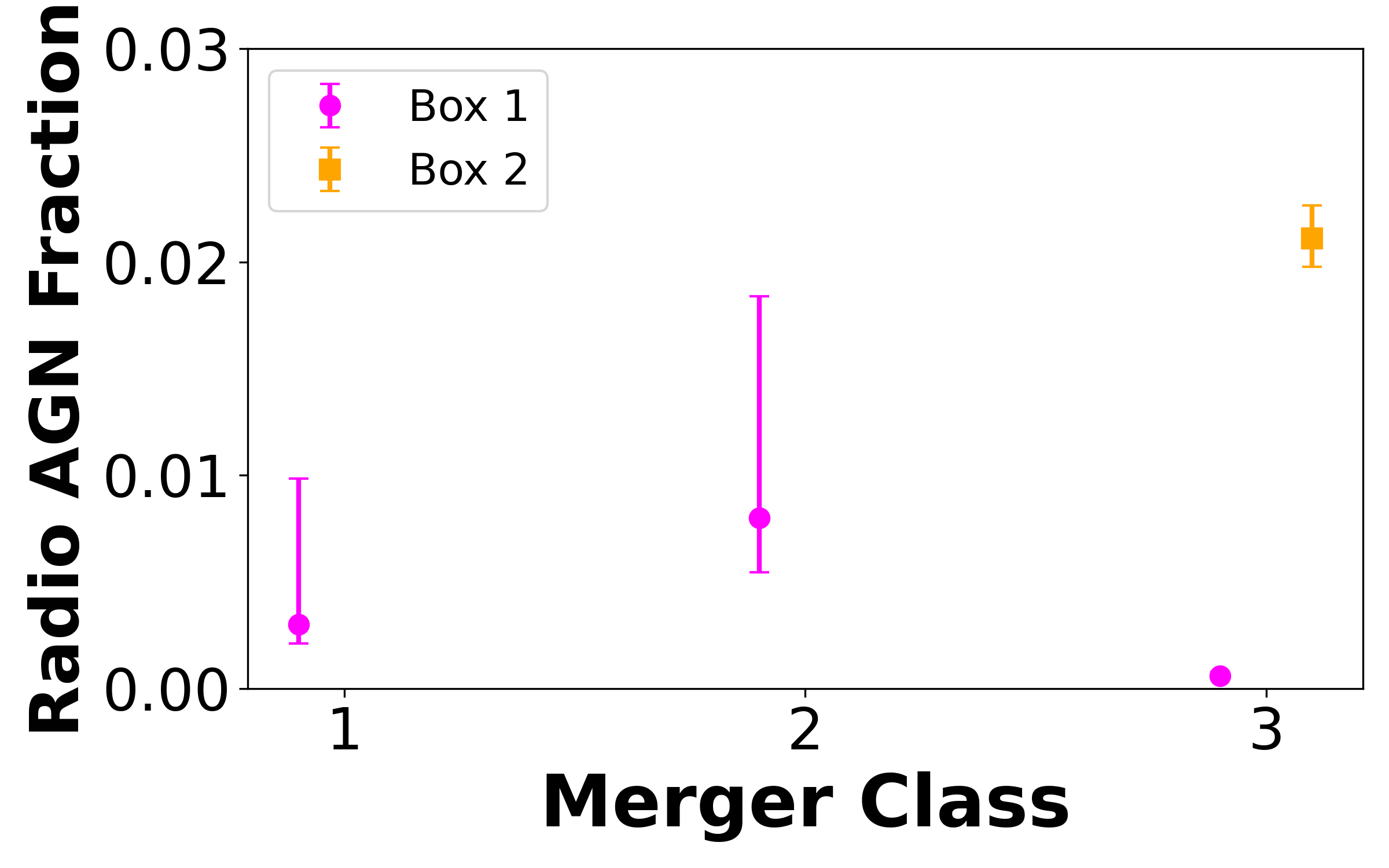}
    \includegraphics[width=0.49\linewidth]{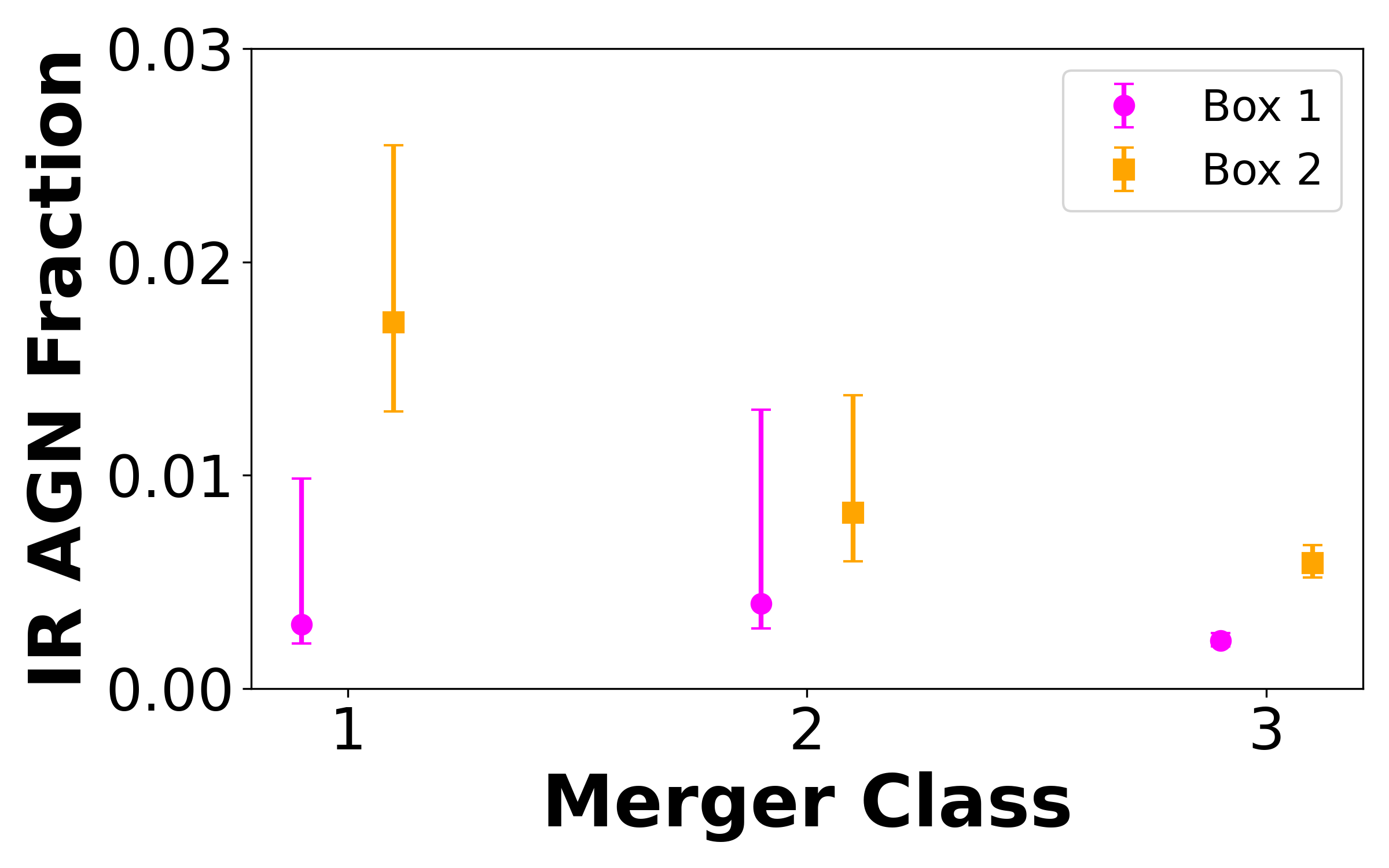}
   \caption{Mean AGN fractions for pairs with $r_p < 50$ kpc separation and all velocity differences, detected in soft X-rays (top left), hard X-rays (top right), radio (bottom left), and infrared (bottom right). The colour scheme and identifications of objects are as in Fig.~\ref{realpairssfr}.}
   \label{realpairsagn}
    \end{figure}

We repeat the same procedure for AGN using all three selection methods in Fig.~\ref{realpairsagn}. However, we have to bundle all pairs with $r_p < 50$ kpc and all velocity differences, due to the relatively smaller number of detected objects. The results are also broadly consistent with AGN fractions found in Fig.~\ref{violinX1}, Fig.~\ref{violinX2}, Fig.~\ref{violinradio}, and Fig.~\ref{violinIR}, suggesting that even in closely interacting pairs that are more likely to merge in the near future, AGN activity is not significantly affected.

\section{Discussion}
\label{discussion}

In this work we have compared SFRs and indicators of AGN activity in samples of progressively closer dynamical galaxy pairs, selected from the unWISE survey to define a mass-selected sample. We compare two absolute magnitude ranges, centered on $M_{W1}=-22.5$ and $-25.5$ and over 1.5 mag either side, to select major mergers with typical magnitude luminosity ratio of 1:3 corresponding to a typical stellar mass ratio 1:4. The centre of these two magnitude ranges correspond to stellar masses of $\log(M_*/M_{\odot}) \sim 10.2$ and $\sim 11.4$ with redshifts of z \(<\) 0.03 and z \(<\) 0.1, respectively.

For both samples, we observe that the star formation activity, as
measured from the $NUV-r$ colour, declines as the galaxies in pairs become
closer in projected separation and velocity difference, and
then rises slightly in the closer bins ($r_p$ \(<\) 50 kpc), mainly for the more massive sample. Unlike some
previous studies \citep[{e.g.,}][]{Patton2013}, we only see this at small projected separations and velocity differences, and not at larger separations. The apparent decline
in star formation (e.g., Fig.~\ref{violin1}) with decreasing
separation is likely not due to interactions (which seem to 
affect morphology only for $r_p < 50$ kpc as in Section \ref{morphology}). When pairs become close enough, evidence for a
relative increase in star formation at $r_p < 20$ kpc is observed in both
mass samples and even for the separate velocity ranges we consider, although more strongly in the more massive sample. An increase in star formation at close 
separations has been detected in several other studies (e.g.,
\citealt{Li2008,Patton2013,Barrera2015,Feng2019,Patton2020}), with claims that the enhancement extends to distances of up to 150 kpc. Here we find that only in the closest separations ($r_p$\( <\) 50 kpc) bin star formation is enhanced, while the fraction of galaxies in pairs that are star-forming actually declines as pair separation decreases. Suppression of star formation in close pairs ( $r_p$ \(<\) 100 kpc) has also been observed by \cite{Feng2024} using SDSS DR7 data, who argue that the suppression in star formation may be due to hydrodynamic effects, including tides, ram stripping from the hot circumgalactic medium and the structure of the pair galaxies. 

While there are some points of agreement with previous work there are also areas where our results differ. We note here that our sample and analysis are different in some respects from previous studies: it is selected by stellar mass using as a proxy data at 3.4 $\mu$m, we measure star formation from $NUV$ colors that are arguably a more reliable and direct measure of star formation over the past $\sim 1$ Gyr \citep{Kaviraj2007}, and we use a wide range of AGN indicators (X-ray and radio fluxes, infrared colours and line ratios in BPT diagrams), yielding similar results. 

Our comparison sample also consists of the wider pairs from the same survey material rather than galaxies selected to be isolated in matched samples (in redshift, stellar mass and environment as in \citealt{Argudo2015,Patton2013}). By construction our samples of galaxies in pairs at increasing separation are matched in redshift and stellar mass (as they are derived from the same material): at $r_p > 200$ kpc these pairs can be considered as non-interacting and effectively isolated (with respect to similarly massive galaxies), while we argue that we likely probe denser environments for closer pairs and that this accounts for the observed trend of quenching as galaxies become closer.

We also note the excellent agreement of our results with the recent work by \cite{Omori2025} who studied close pairs in the Hyper Suprime Cam Subaru Strategic Program to investigate the merger-SFR-AGN connection in galaxy pairs and found little difference in AGN and star formation activity between close pair galaxies and isolated galaxies. It is only among objects that are classified as post-mergers that any sign of significantly increased activity is observed in agreement with \cite{Ellison2024,Ellison2025}.

Simulations (e.g., \citealt{Mihos1996,Hopkins2008,Scudder2012,Moreno2019}) show that interactions and mergers cause strong gravitational torques, possibly resulting in nuclear starbursts \citep{Renaud2015,Moreno2021} and the triggering of AGN. As already mentioned above, while some observations have claimed that the fraction of AGN increases in close galaxy pairs
\citep{Ellison2011,Ellison2013,Satyapal2014,Weston2017}, others
have not supported this claim \citep{Shah2020,Jin2021}. We find no strong evidence that AGN activity
increases as a function of pair separation. This is consistent with the hypothesis that, at least at lower redshifts, secular processes may be more important than mergers in triggering AGN activity (e.g., \citealt{Menci2014,Mahoro2017,Mahoro2019,Hickox2018}).
Even in the simulations by \cite{McAlpine2020} the AGN enhancement in pairs is modest. Most of the star formation and AGN activity may occur post-coalescence \citep{Ellison2024,Ellison2025} rather than during the early interaction phase (e.g., see \citealt{Omori2025}. However, post-starburst galaxies (usually considered as merger or interaction remnants) do not generally host powerful AGN \citep{Nielsen2012,DePropris2014b,
Lanz2022,Almaini2025}.

\begin{figure}
    \centering
    \includegraphics[width=1\linewidth]{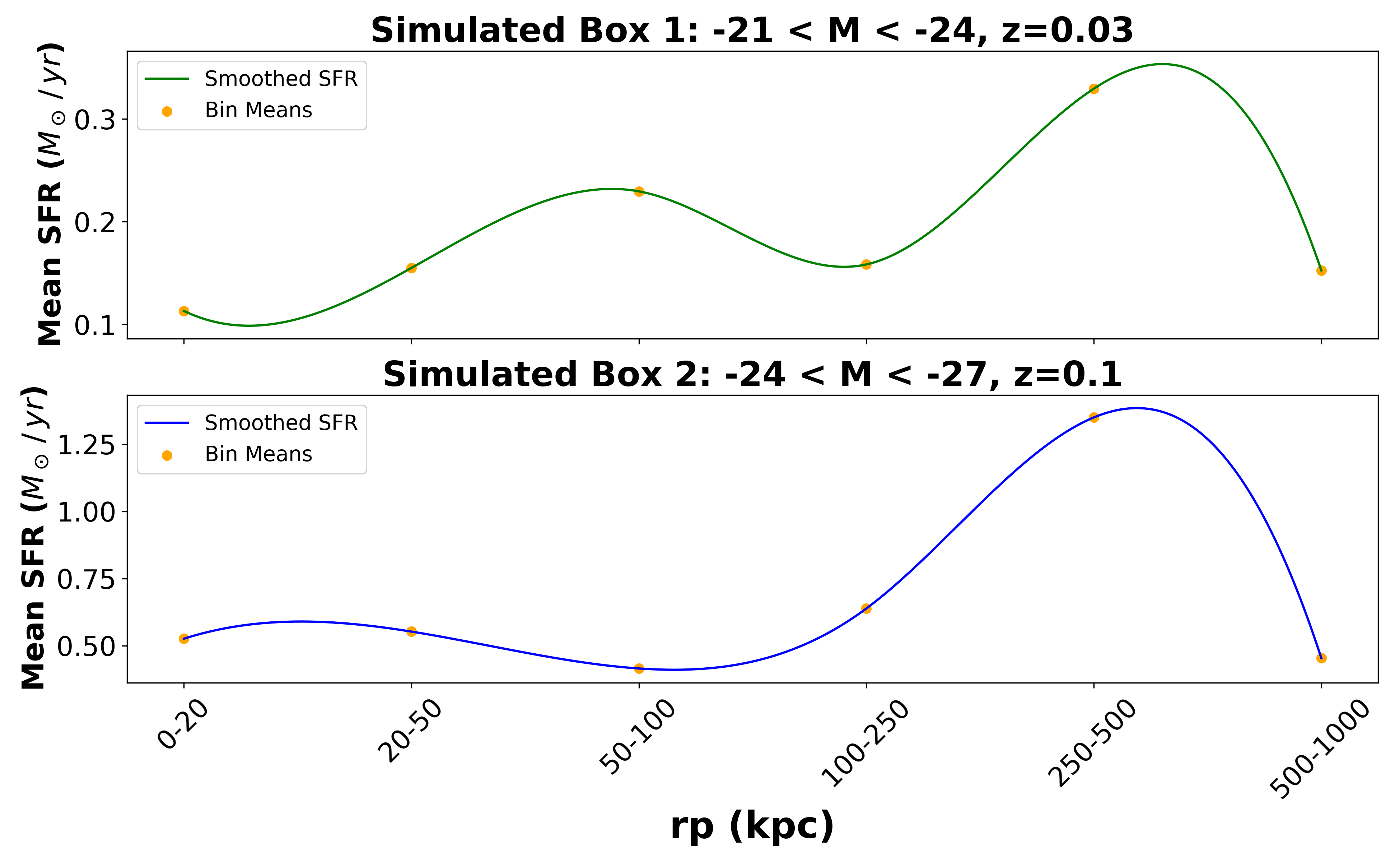}
    \caption{Smoothed SFRs for galaxies in pairs from the Illustris-TNG simulation versus 3D separation for simulated Box 1 (top) and Box 2 (bottom).}
    \label{TNGSFR}
\end{figure}

\begin{figure}
    \centering
    \includegraphics[width=1\linewidth]{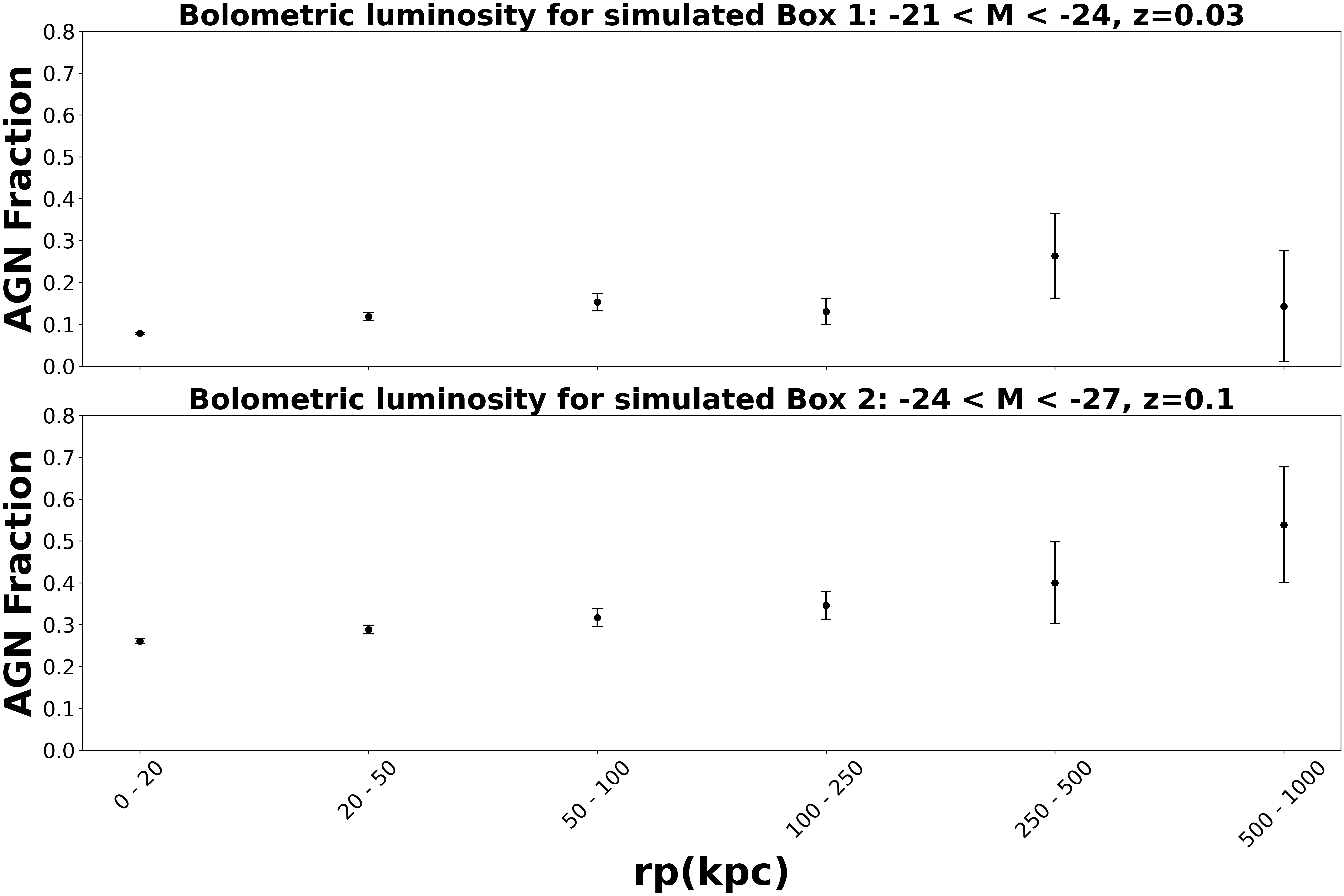}
    \caption{Bolometric luminosity of luminous AGNs (\(L_{Bol} > 10^{44}\) erg/s) for galaxies in pairs from the Illustris-TNG simulation versus 3D separation for simulated Box 1 (top) and Box 2 (bottom).}
    \label{TNGAGN}
\end{figure}

As a test of our hypothesis, we have extracted a sample of pairs of similar masses from the $z=0.1$ snapshot of the
Illustris-TNG300 simulations \citep{Pillepich2018}. First, we notice that the simulation identifies more pairs at closer separation than what we obtain from observations, a result also observed by \cite{Faria2025}. We plot star formation and bolometric AGN luminosity as a function of 3D pair separation in Fig.~\ref{TNGSFR} and Fig.~\ref{TNGAGN}, respectively.

Star formation is observed to decline as a function of pair separation, albeit more sharply than we observe. However, we see no strong enhancement in star formation at closer separation, while we observe a modest increase at $r_p < 20$ kpc. This is different in some respects from previous work and we suggest that selection by stellar mass and the use of the $NUV$ colours to assess star formation plays at least some role in our results.

In agreement with our observations, there is little or no AGN enhancement as a function of decreasing separation between galaxies in pairs. This is consistent with some simulations, while others claim a stronger effect (e.g., \citealt{Steinborn2016}).

\section{Conclusions}
\label{conclusion}

We have selected a large sample of galaxy pairs with six ranges of projected separations within $0 < r_p < 1000$ kpc and two velocity ranges within $0 < \Delta V < 1000$ km
s$^{-1}$ from the unWISE survey \citep{Lang2014,Schlafly2019}
and with redshifts of \(z < 0.25\) from the NED-LVS compilation of \cite{Cook2023}. We find 12,122 pairs with luminosity ratio 
of 1:3 at $M_{W1} = -22.5 \pm 1.5$ mag and 39,757 pairs at
$M_{W1} = -25.5 \pm 1.5$ with z \(<\) 0.03 and z \(<\) 0.10, respectively. We compare the fractions of star-forming galaxies in pairs and AGN activity as a function of pair separation (in a relative sense, rather than absolute values).

Comparing star formation as a function of pair separation and velocity differences in Figure \ref{violin1}, we find evidence for a decline in star formation as galaxies in pairs become closer, probably due to environmental effects as there is no evidence of interactions at $r_p > 50$ kpc that may cause quenching, and then a modest increase at closer separations (i.e., $r_p \lesssim 50$ kpc) likely due to the interaction between the pair members.

We find no strong evidence for an increase in AGN activity when AGN are selected by different methods (X-ray, radio, infrared, and optical) as shown in Figures \ref{violinX1}, \ref{violinX2}, \ref{violinradio}, \ref{violinIR}, and \ref{BPT1}, suggesting that galaxy interactions do not highly enhance AGN activity and may be triggered by secular processes in the local universe.\\

\begin{acknowledgements}
We thank the anonymous referees for accepting to review this paper and for providing constructive and valuable comments that improved the manuscript.We acknowledge Emmanuel Bempong from the University of Manchester, and Betelehem Bilata and Jaime Perea from the IAA-CSIC for the assistance in running the part of the code that required supercomputers. JC and RdP acknowledge the financial support from the Education, Audio and Cultural Executive Agency of the European Commission through the Pan-African Planetary and Space Science Network under funding agreement number 62421.24-PANAF-12020-1-BW-PANAF-MOBAF. MP acknowledges the support from the Spanish Ministerio de Ciencia e
Innovación-Agencia Estatal de Investigación through projects PID2022-140871NB-C21 and PID2024-162972NB-I00, the State Agency for Research of the Spanish MCIU through the Center of Excellence Severo Ochoa award to the Instituto de Astrofísica de Andalucía (CEX2021-001131-S funded by MCIN/AEI/10.13039/501100011033), and the Space Science and Geospatial Institute
(SSGI) under the Ethiopian Ministry of Innovation and Technology (MInT).\\ This publication makes use of data products from the Wide-field Infrared Survey Explorer, which is a joint project of the University of California, Los Angeles, and the Jet Propulsion Laboratory/California Institute of Technology, funded by the National Aeronautics and Space Administration.\\
This research also made use of the NASA/IPAC Extragalactic Database (NED),
which is operated by the Jet Propulsion Laboratory, California Institute of Technology, under contract with the National Aeronautics and Space Administration.
\\
The Pan-STARRS1 Surveys (PS1) and the PS1 public science archive have been made possible through contributions by the Institute for Astronomy, the University of Hawaii, the Pan-STARRS Project Office, the Max-Planck Society and its participating institutes, The Johns Hopkins University, Durham University, the University of Edinburgh, the Queen's University Belfast, the Harvard-Smithsonian Center for Astrophysics, the Las Cumbres Observatory Global Telescope Network Incorporated, the National Central University of Taiwan, the Space Telescope Science Institute, the National Aeronautics and Space Administration under Grant No. NNX08AR22G issued through the Planetary Science Division of the NASA Science Mission Directorate, the National Science Foundation Grant No. AST–1238877, the University of Maryland, Eotvos Lorand University (ELTE), the Los Alamos National Laboratory, and the Gordon and Betty Moore Foundation.
\\
The National Radio Astronomy Observatory is a facility of the National Science Foundation operated under cooperative agreement by Associated Universities, Inc.
\\
The national facility capability for SkyMapper has been funded through ARC LIEF grant LE130100104 from the Australian Research Council, awarded to the University of Sydney, the Australian National University, Swinburne University of Technology, the University of Queensland, the University of Western Australia, the University of Melbourne, Curtin University of Technology, Monash University and the Australian Astronomical Observatory. SkyMapper is owned and operated by The Australian National University's Research School of Astronomy and Astrophysics. The survey data were processed and provided by the SkyMapper Team at ANU. The SkyMapper node of the All-Sky Virtual Observatory (ASVO) is hosted at the National Computational Infrastructure (NCI). Development and support of the SkyMapper node of the ASVO has been funded in part by Astronomy Australia Limited (AAL) and the Australian Government through the Commonwealth's Education Investment Fund (EIF) and National Collaborative Research Infrastructure Strategy (NCRIS), particularly the National eResearch Collaboration Tools and Resources (NeCTAR) and the Australian National Data Service Projects (ANDS).
\\
This work also used data from eROSITA, the X-ray instrument aboard SRG, a joint Russian-German science mission supported by the Russian Space Agency (Roskosmos), in the interests of the Russian Academy of Sciences represented by its Space Research Institute (IKI), and the Deutsches Zentrum für Luft- und Raumfahrt (DLR). The development and construction of the eROSITA X-ray instrument was led by MPE, with contributions from the Dr. Karl Remeis Observatory Bamberg \& ECAP (FAU Erlangen-Nuernberg), the University of Hamburg Observatory, the Leibniz Institute for Astrophysics Potsdam (AIP), and the Institute for Astronomy and Astrophysics of the University of Tübingen, with the support of DLR and the Max Planck Society. The Argelander Institute for Astronomy of the University of Bonn and the Ludwig Maximilians Universität Munich also participated in the science preparation for eROSITA.
\\
GALEX is operated for NASA by the California Institute of Technology under NASA contract NAS5-98034.
\end{acknowledgements}

\bibliographystyle{aa} 
\bibliography{aa} 
\appendix
\section{Additional Tables}
\label{appendix}

\begin{table}[ht!]
\centering
 \caption{Fraction of soft X-ray selected AGN in pairs per separation and velocity difference.}
    \label{table:x_ray_soft}
\begin{tabular}{lcccc}
  \hline
     $r_p$   & \textbf{Box 1 }&\textbf{Box 1 } & \textbf{Box 2 } & \textbf{Box 2} \\
  \hline
      kpc & ($\Delta V < 500$) km s$^{-1}$ & ($500 < \Delta V < 1000$) km s$^{-1}$ & ($\Delta V < 500$) km s$^{-1}$ & ($500 < \Delta V < 1000$) km s$^{-1}$ \\
      \hline
    \hline
    $<20$ &  $0.056^{+0.065}_{-0.018}$ & $0.500^{+0.203}_{-0.203}$ & $0.200^{+0.101}_{-0.057}$ & $0.600^{+0.157}_{-0.215}$ \\ 
    $20 < r_p < 50$ & $0.060^{+0.043}_{-0.018}$ & $0.000^{+0.231}_{-0.024}$ & $0.426^{+0.065}_{-0.060}$ & $0.579^{+0.100}_{-0.115}$ \\
    $50 < r_p < 100$ & $0.012^{+0.027}_{-0.004}$ & $0.182^{+0.164}_{-0.065}$ & $0.460^{+0.045}_{-0.044}$ & $0.457^{+0.084}_{-0.080}$ \\
    $100 < r_p < 250$ & $0.022^{+0.020}_{-0.007}$ & $0.040^{+0.081}_{-0.013}$ & $0.378^{+0.027}_{-0.026}$ & $0.563^{+0.043}_{-0.044}$ \\ 
   $250 < r_p < 500$& $0.024^{+0.023}_{-0.008}$ & $0.000^{+0.054}_{-0.005}$ & $0.442^{+0.026}_{-0.026}$ & $0.496^{+0.042}_{-0.042}$ \\
    $500 < r_p < 1000$ & $0.026^{+0.020}_{-0.007}$ & $0.000^{+0.046}_{-0.004}$ & $0.409^{+0.023}_{-0.022}$ & $0.381^{+0.037}_{-0.035}$ \\
    \hline
\end{tabular}
\end{table}

\begin{table}[ht!]
\centering
\caption{Fraction of hard X-ray selected AGN in pairs per separation and velocity difference.}
    \label{table:x_ray_hard}
\begin{tabular}{lcccc}
  \hline
     $r_p$   & \textbf{Box 1 }&\textbf{Box 1 } & \textbf{Box 2 } & \textbf{Box 2} \\
  \hline
      kpc & ($\Delta V < 500$) km s$^{-1}$ & ($500 \leq \Delta V < 1000$) km s$^{-1}$ & ($\Delta V < 500$) km s$^{-1}$ & ($500 \leq \Delta V < 1000$) km s$^{-1}$ \\
      \hline
    \hline
    $<20$ &  $0.111^{+0.116}_{-0.038}$ & $0.000^{+0.459}_{-0.056}$ & $0.400^{+0.159}_{-0.125}$ & $0.750^{+0.103}_{-0.274}$ \\ 
    $20 < r_p < 50$ & $0.071^{+0.062}_{-0.022}$ & ---* & $0.600^{+0.075}_{-0.086}$ & $0.733^{+0.082}_{-0.137}$ \\
    $50 < r_p < 100$ & $0.024^{+0.051}_{-0.007}$ & $0.250^{+0.274}_{-0.103}$ & $0.560^{+0.055}_{-0.058}$ & $0.619^{+0.091}_{-0.112}$ \\
    $100 < r_p < 250$ & $0.037^{+0.034}_{-0.011}$ & $0.071^{+0.132}_{-0.024}$ & $0.542^{+0.038}_{-0.039}$ & $0.556^{+0.053}_{-0.056}$ \\ 
   $250 < r_p < 500$& $0.015^{+0.033}_{-0.004}$ & $0.000^{+0.123}_{-0.012}$ & $0.518^{+0.035}_{-0.036}$ & $0.483^{+0.053}_{-0.053}$ \\
    $500 < r_p < 1000$ & $0.045^{+0.033}_{-0.013}$ & $0.000^{+0.064}_{-0.006}$ & $0.565^{+0.030}_{-0.030}$ & $0.495^{+0.049}_{-0.049}$ \\
    \hline
\end{tabular}
\small{*Refers to the bin whose total number of galaxies in pairs is less than 4, as a result was excluded from the analysis.}
\end{table}

\begin{table}[ht!]
\centering
\caption{Fraction of radio-selected AGN in pairs and velocity difference.}
    \label{table:radio}
\begin{tabular}{lcccc}
  \hline
     $r_p$   & \textbf{Box 1 }&\textbf{Box 1 } & \textbf{Box 2 } & \textbf{Box 2} \\
  \hline
      kpc & ($\Delta V < 500$) km s$^{-1}$ & ($500 < \Delta V < 1000$) km s$^{-1}$ & ($\Delta V < 500$) km s$^{-1}$ & ($500 < \Delta V < 1000$) km s$^{-1}$ \\
      \hline
    \hline
    $<20$ &  $0.039^{+0.036}_{-0.012}$ & $0.000^{+0.264}_{-0.028}$ & $0.400^{+0.061}_{-0.055}$ & $0.250^{+0.155}_{-0.083}$ \\ 
    $20 < r_p < 50$ & $0.027^{+0.017}_{-0.007}$ & $0.000^{+0.142}_{-0.014}$ & $0.465^{+0.042}_{-0.041}$ & $0.571^{+0.078}_{-0.085}$ \\
    $50 < r_p < 100$ & $0.007^{+0.010}_{-0.002}$ & $0.059^{+0.113}_{-0.019}$ & $0.418^{+0.032}_{-0.031}$ & $0.673^{+0.058}_{-0.070}$ \\
    $100 < r_p < 250$ & $0.000^{+0.004}_{-0.000}$ & $0.000^{+0.023}_{-0.002}$ & $0.422^{+0.020}_{-0.020}$ & $0.536^{+0.035}_{-0.036}$ \\ 
   $250 < r_p < 500$& $0.006^{+0.006}_{-0.002}$ & $0.009^{+0.019}_{-0.003}$ & $0.403^{+0.019}_{-0.018}$ & $0.444^{+0.033}_{-0.032}$ \\
    $500 < r_p < 1000$ & $0.002^{+0.004}_{-0.000}$ & $0.007^{+0.016}_{-0.002}$ & $0.332^{+0.016}_{-0.015}$ & $0.306^{+0.025}_{-0.023}$ \\
    \hline
\end{tabular}
\end{table}

\begin{table}[ht!]
\centering
\caption{Fraction of infrared selected AGN in pairs per separation and velocity difference.}
    \label{table:ir}
\begin{tabular}{lcccc}
  \hline
     $r_p$   & \textbf{Box 1 }&\textbf{Box 1 } & \textbf{Box 2 } & \textbf{Box 2} \\
  \hline
      kpc & ($\Delta V < 500$) km s$^{-1}$ & ($500 < \Delta V < 1000$) km s$^{-1}$ & ($\Delta V < 500$) km s$^{-1}$ & ($500 < \Delta V < 1000$) km s$^{-1}$ \\
      \hline
    \hline
    $<20$ &  $0.0049^{+0.0048}_{-0.0015}$ & $0.000^{+0.0419}_{-0.0040}$ & $0.0150^{+0.0073}_{-0.0036}$ & $0.000^{+0.0144}_{-0.0014}$ \\ 
    $20 < r_p < 50$ & $0.0049^{+0.0026}_{-0.0012}$ & $0.000^{+0.0082}_{-0.0008}$ & $0.0103^{+0.0027}_{-0.0018}$ & $0.0016^{+0.0036}_{-0.0005}$ \\
    $50 < r_p < 100$ & $0.0026^{+0.0015}_{-0.0007}$ & $0.000^{+0.0045}_{-0.0004}$ & $0.0056^{+0.0015}_{-0.0010}$ & $0.0040^{+0.0027}_{-0.0011}$ \\
    $100 < r_p < 250$ & $0.0016^{+0.0008}_{-0.0004}$ & $0.0009^{+0.0021}_{-0.0003}$ & $0.0036^{+0.0006}_{-0.0005}$ & $0.0031^{+0.0012}_{-0.0007}$ \\ 
   $250 < r_p < 500$& $0.0027^{+0.0010}_{-0.0006}$ & $0.0018^{+0.0024}_{-0.0006}$ & $0.0041^{+0.0006}_{-0.0004}$ & $0.0043^{+0.0010}_{-0.0007}$ \\
    $500 < r_p < 1000$ & $0.0026^{+0.0009}_{-0.0005}$ & $0.0007^{+0.0016}_{-0.0002}$ & $0.0052^{+0.0005}_{-0.0004}$ & $0.0060^{+0.0009}_{-0.0007}$ \\
    \hline
\end{tabular}
\end{table}
\end{document}